\documentclass[journal]{IEEEtran}
\usepackage[OT1]{fontenc}
\usepackage[utf8]{inputenc}
\usepackage{amsmath}
\usepackage{amssymb}
\usepackage{graphicx}

\makeatletter
\IEEEoverridecommandlockouts

\usepackage{color}
\usepackage{bm}
\usepackage{algorithm}
\usepackage{algorithmic}
\usepackage{multirow}
\usepackage{booktabs}
\usepackage{array}
\usepackage{amsthm}
\usepackage{lipsum}
\usepackage{enumerate}
\usepackage{stfloats}
\usepackage{subfigure}
\usepackage{cases}
\usepackage{diagbox}
\usepackage{threeparttable}
\usepackage{cite}
\usepackage{url}

\usepackage{cuted}
\usepackage{stfloats}
\usepackage{bm}
\usepackage{graphicx}
\usepackage{makecell}
\usepackage{threeparttable}

\newtheorem{remark}{Remark}
\newcommand{\non}{\nonumber}

\begin{document}
	
	\title{Active Beyond-Diagonal Reconfigurable \\ Intelligent Surface with Hybrid \\
		Transmitting and Reflecting Mode }
	\author{Fu Liu, \IEEEmembership{Graduate Student Member,~IEEE}, Hongyu Li, \IEEEmembership{Member,~IEEE}, and Shanpu Shen,~\IEEEmembership{Senior Member,~IEEE}
		\thanks{Manuscript received; The work of H. Li is funded by the National Natural Science Foundation of China (grant no. 62501509) and the Natural Science Foundation of Guangdong Province (grant no. 2026A1515011048). The work of S. Shen is funded by the Science and Technology Development Fund, Macau SAR (File/Project no. 001/2024/SKL), by University of Macau (File no. SRG2025-00060-IOTSC), and by University of Macau Development Foundation (UMDF) (File no. UMDF-TISF-I/2026/025/IOTSC). \textit{(Corresponding author: Hongyu Li.)}}
		\thanks{F. Liu and H. Li are with the Internet of Things Thrust, The Hong Kong University
			of Science and Technology (Guangzhou), Guangzhou 511400, China (e-mail: fliu137@connect.hkust-gz.edu.cn; hongyuli@hkust-gz.edu.cn).}
		\thanks{S. Shen is with the State Key Laboratory of Internet of Things for Smart
			City and Department of Electrical and Computer Engineering, University of
			Macau, Macau, China (e-mail:shanpushen@um.edu.mo).}}
	
	\maketitle
	\begin{abstract}
		
		Beyond-diagonal reconfigurable intelligent surfaces (BD-RISs), originally in the passive form, have attracted attention due to their benefits in enhanced wave manipulating through flexible inter-element connections and element arrangements. To mitigate the severe multiplicative fading, the concept of active BD-RISs with signal amplification capability has recently been proposed.   
        Inspired by this, we investigate the hybrid transmitting and reflecting mode of active BD-RISs to achieve full-space coverage.
		We start by deriving a physics compliant communication model applying active BD-RIS with hybrid mode. 
		We further propose novel architectures including reciprocal and non-reciprocal implementations with cell-wise single, group, and fully connections. 
		We also develop a unified optimization framework for the joint transmit precoding and hybrid mode active BD-RIS design to maximize the sum rate of multi-user communication systems, which is applicable to all considered architectures.
		Numerical results demonstrate that, under the same total power budget, the proposed active BD-RIS with hybrid mode substantially outperforms active and passive simultaneous transmitting and reflecting RISs as well as passive BD-RISs with hybrid mode. This shows the synergy gain from inter-element connection, element arrangements, and active amplification.

	\end{abstract}
	

	\section{Introduction}
	\label{sec:intro}
	
	Beyond-diagonal reconfigurable intelligent surfaces (BD-RISs) represent an emerging and general framework for RIS technologies \cite{SGong2019,K-KWong,di2020smart,wu2021intelligent}.  
	The key difference between BD-RIS and conventional diagonal (D) RIS that has diagonal phase shift matrices is that BD-RIS generates scattering matrices that are not limited to being diagonal \cite{nerini2023closed}. This is achieved by introducing inter-element connections, which allow electromagnetic waves to propagate from one element to another \cite{bdristutorial}.
	Such inter-element interactions introduce additional degrees of freedom for wave manipulation, thereby enhancing system performance from various perspectives \cite{khan2024beyond}.

\begin{figure*}
		\centering \includegraphics[height=2.5 in]{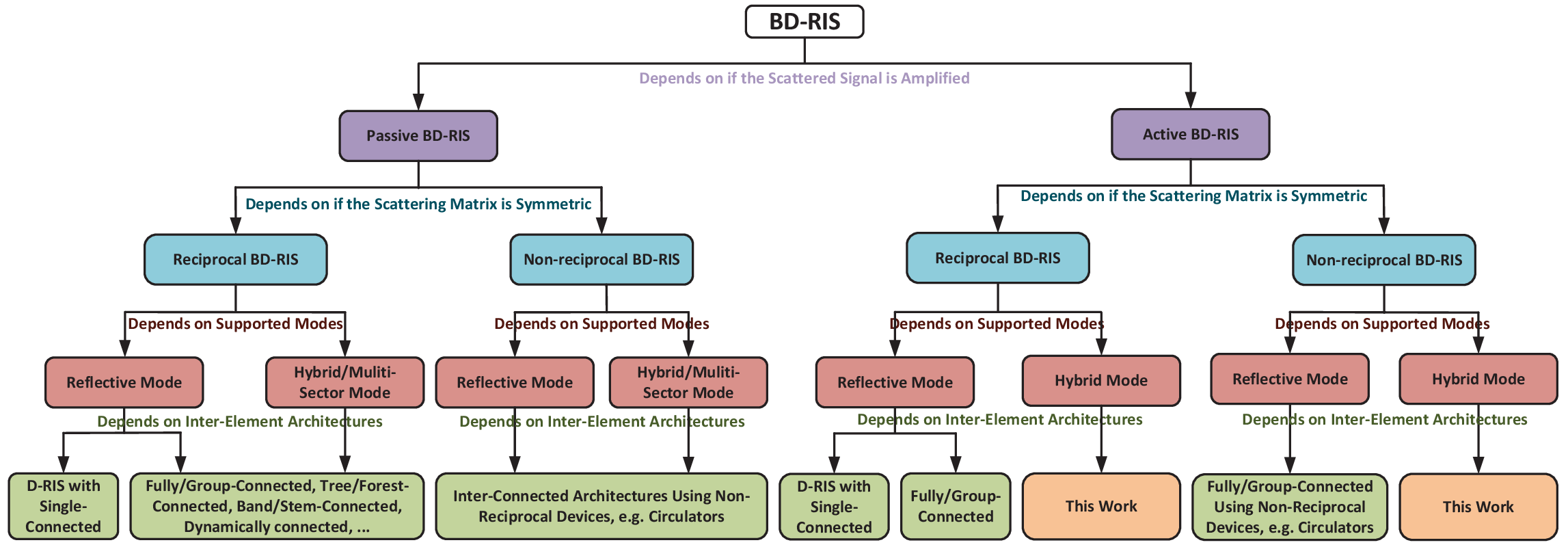}
		\caption{BD-RIS classification tree.}
		\label{fig:RIS_tree}
	\end{figure*}

	BD-RIS, in its passive form where scattered signals are not amplified, has been extensively investigated in modeling and architecture design \cite{SShen,graph_theory,bandconnect,zhou2025novel,nonreRIS,first,multisector}, signal processing \cite{santamaria2024mimo,zhou2025joint,Li2024a}, hardware impairments \cite{Nerini2025,lossy2026,Nerini2024d}, and system-level applications \cite{ISAC,WPT}. 
	From the modeling and architecture design perspective, passive BD-RIS can be classified according to the left branch of the classification tree in Fig. \ref{fig:RIS_tree}. Specifically, group and fully-connected architectures implemented by reciprocal reconfigurable impedance networks was first proposed in \cite{SShen}.
	To balance performance gains and circuit complexity, tree and forest-connected architectures were subsequently proposed in \cite{graph_theory} using graph-theoretic modeling, aiming at achieving the performance upper bound for BD-RIS aided single-antenna systems with the least circuit complexity. 
	This result has been recently extended to multi-antenna multi-user systems where the hardware-efficient band- and stem-connected BD-RIS architecture can achieve the performance upper bound \cite{bandconnect,zhou2025novel}.
In addition to reciprocal architectures, non-reciprocal architectures implemented using devices such as circulators and isolators have also been investigated in \cite{nonreRIS}, with the aim to generate asymmetric beams for downlink and uplink, so as to benefit full-duplex and physical layer secrecy scenarios. 
	On the other hand, unlike D-RIS whose reflection functionality is limited to signals impinging on one side of the surface, BD-RIS has been proven to support more advanced hybrid transmitting and reflecting mode \cite{first} and multi-sector mode \cite{multisector} with full-space wireless coverage.
	From the signal processing perspective, beamforming design \cite{santamaria2024mimo,zhou2025joint} and channel estimation \cite{Li2024a} for BD-RIS aided communication systems have been studied.
	From the hardware impairments perspective, practical hardware effects such as lossy interconnections \cite{Nerini2025,lossy2026} and mutual coupling \cite{Nerini2024d} have been modeled and analyzed, revealing their impacts on system performance.
	In addition to these aspects, passive BD-RIS has also been explored in various scenarios, such as integrated sensing and communication (ISAC) \cite{ISAC} and wireless power transfer (WPT) \cite{WPT}.
	Despite aforementioned advances,
    the performance enhancement achieved by passive BD-RIS is still limited due to the multiplicative fading effect, where the cascaded path loss scales with the product of the transmitter-RIS and RIS-receiver path losses \cite{ruize2021}. 
	In this sense, passive RISs are more useful in scenarios where the line-of-sight transmission links are blocked by obstacles, while their benefits in scenarios with line-of-sight links might be marginal.
    
    To overcome this limitation, the concept of active RISs that use active devices to amplify scattered signals has been proposed.
	In the family of active RISs, active D-RIS with a diagonal scattering matrix whose diagonal elements have controllable phase shifts and amplitudes has been first proposed in \cite{prevail2023}. Specifically, the amplitude of the scattered signals can be magnified by reflection-type amplifiers to compensate for the multiplicative path loss and enhance system performance. 
	Building upon this architecture, extensive efforts have been devoted to active D-RIS aided wireless communications, including transmission design \cite{partialCSI2024,allu2023robust,Yaswanth2024,Chen2022a,Ye2025}, architecture design \cite{Zhu2023a}, and performance analysis under hardware impairments \cite{Yue2024,Peng2024}.
	With the ability to jointly adjust the phase shift and amplitude gain, active D-RIS has been 
    employed to improve the performance of various wireless systems such as symbiotic radio \cite{Lyu2023}, ISAC \cite{ming2024} and simultaneous wireless information and power transfer (SWIPT) \cite{Zhai2024}. 
	Note that active D-RIS still operates in a reflection-only mode, which provides limited coverage for users located on the opposite side of the surface.
	To address this issue, active simultaneously transmitting and reflecting (STAR) RIS \cite{xu2023,Maghrebi2024,Faramarzi2025,wang2026,aung2025,rongguang2025,huang2025}, also referred to as active double-faced RIS in some studies \cite{Liu2022c,Guo2023,Zhou2023a}, has recently emerged as a promising technology, enabling the surface to achieve full space coverage while providing amplification capabilities.  
	Active STAR-RIS aided communication systems have been investigated from perspectives of modeling and fundamental performance analysis \cite{xu2023} as well as system-level design \cite{Liu2022c,Guo2023,Zhou2023a,Maghrebi2024}. 
	In addition to communication scenarios, active STAR-RIS has also been explored in several emerging wireless applications, 
	such as SWIPT \cite{Faramarzi2025}, wireless powered communications \cite{wang2026}, and edge computing \cite{aung2025}. 
	Meanwhile, practical issues such as hardware impairments and prototype implementation have also begun to attract attention in recent works \cite{huang2025,rongguang2025}.

	The aforementioned studies mainly focus on active D-RISs or active STAR-RISs, both of which are modeled by directly introducing an amplification factor to each element with additive dynamic noise.
	However, such modeling fails to capture the intrinsic electromagnetic characteristics introduced by the interaction between reflection-type amplifiers and reconfigurable impedance components.
    To establish a physically compliant model for active RISs, a rigorous model for active RIS aided communication systems was first developed based on multiport network analysis and reciprocal or non-reciprocal group/fully-connected architectures were subsequently proposed in \cite{shen2026active}, as summarized in the right branch of the BD-RIS classification tree in Fig. \ref{fig:RIS_tree}. 
	While \cite{shen2026active} primarily focused on active BD-RIS with reflecting mode, it opens the door to 
    explore further extension to different modes of active BD-RIS that can support to enlarge wireless coverage.
    Motivated by this, in this work, we develop a comprehensive and physically-compliant modeling framework
	for active BD-RIS with hybrid transmitting and reflecting mode, as highlighted in Fig. \ref{fig:RIS_tree}. 
    Specifically, the contributions can be summarized as follows.

	\textit{First}, we derive and rigorously analyze a wireless communication model for active BD-RIS with hybrid transmitting and reflecting mode.
	Specifically, the proposed model characterizes the hybrid mode active BD-RIS as a multiport passive reconfigurable impedance network interconnected with multiple back to back placed antennas and reflection-type amplifiers, thereby explicitly revealing the effect of signal flow, electromagnetic coupling, and amplification.

	\textit{Second}, we propose novel cell-wise single, group and fully-connected architectures based on reciprocal and non-reciprocal circuit designs and rigorously characterize the intrinsic constraints. We also prove that active STAR-RIS is a special case of the proposed reciprocal hybrid mode active BD-RIS with cell-wise single-connected architecture.

	\textit{Third}, we formulate and solve a joint precoding and hybrid mode active BD-RIS design problem for a multi-user multiple input single output (MU-MISO) system with sum-rate maximization as the objective..
    To address this challenging nonconvex problem, we develop a unified optimization framework that is applicable to hybrid mode active BD-RIS with cell-wise single/group/fully-connected architectures. We further present a comprehensive analysis of the computational complexity of the proposed unified optimization algorithm.
	
	\textit{Fourth}, we present simulation results to evaluate the sum rate performance of the MU-MISO system aided by hybrid mode active BD-RIS. Simulation results show that, under the identical total power budget, the proposed hybrid mode active BD-RIS substantially outperforms active and passive STAR-RIS as well as passive BD-RIS with hybrid mode.
	
	\textit{Organization:} Section \ref{sec:model} provides a communication model aided by hybrid mode active BD-RIS. 
    Section \ref{sec:architecture} proposes 
    reciprocal and non-reciprocal hybrid mode active BD-RIS with cell-wise single, group, and fully-connected architectures. Section \ref{sec:joint optimization} provides a unified optimization algorithm for a communication system aided by the hybrid mode active BD-RIS. Section \ref{sec:evaluation} provides the performance evaluation of the proposed algorithm. Section \ref{sec:conclusion} concludes this paper.
	
    \textit{Notation:} Boldface lower-case letters denote vectors. Boldface upper-case letters denote matrices. $(\cdot)^{\mathsf{T}}$, $(\cdot)^{*}$, $(\cdot)^{\mathsf{H}}$, and $(\cdot)^{-1}$ denote transpose, conjugate, Hermitian transpose, and inverse, respectively. $\mathbb{C}$ and $\mathbb{R}$ are the complex and real fields. $\mathbb{E}[\cdot]$ denotes expectation. $\Re\{\cdot\}$ denotes the real part. 
    $\mathsf{diag}(\cdot)$ and $\mathsf{blkdiag}(\cdot)$ denote diagonal matrices and block diagonal matrices, respectively.
    $\|\cdot\|_{\mathsf{F}}$ and $\|\cdot\|_{2}$ denote the Frobenius norm and Euclidean $\ell_{2}$ norm, respectively. $\jmath=\sqrt{-1}$ is the imaginary unit.
    $\mathbf{I}_N$ and $\mathbf{0}$ denote the $N\times N$ identity matrix and an all-zero matrix with proper dimensions, respectively.
    $a\sim\mathcal{CN}(0,\sigma^{2})$ denotes a zero-mean circularly symmetric complex Gaussian random variable with variance $\sigma^{2}$.
    $[\mathbf{A}]_{i:i',j:j'}$ denotes the submatrix of $\mathbf{A}$ formed by rows $i$ to $i'$ and columns $j$ to $j'$. $\otimes$ denotes the Kronecker product.
    $\mathrm{vec}(\cdot)$ and $\mathrm{Tr}(\cdot)$ denote the vectorization and trace of a matrix, respectively.
	
	\section{Active BD-RIS Aided Communication Model}
	\label{sec:model}
\begin{figure}
		\centering \includegraphics[width=0.48\textwidth]{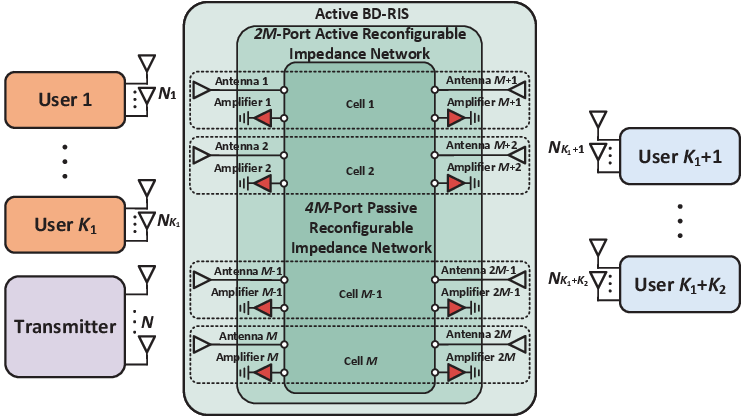}
		\caption{Diagram of an $M$-cell hybrid mode active BD-RIS aided communication system.}
		\label{fig:RIS_NA}
	\end{figure}

	Consider an active BD-RIS aided wireless communication system, where a transmitter having $N_\mathrm{T}$ antennas indexed by $\mathcal{N}_\mathrm{T}=\{1,\ldots,N_\mathrm{T}\}$
	serves $K$ multi-antenna users with the assistance of a $2M$-element active BD-RIS. 
	User $k$ has $N_{\mathrm{R},k}$ antennas, $\forall k\in\mathcal{K}=\{1,\ldots,K\}$.
	The active BD-RIS is constructed by $M$ cells\footnote{The concept of ``cell''  was first introduced in \cite{first}, indicating that two elements are connected with each other via reconfigurable impedance network.}, indexed by $\mathcal{M}=\{1,\ldots,M\}$, where the
	$n$th cell consists of element $n$ and element $n+M$, as illustrated in Fig. \ref{fig:RIS_NA}. Therefore, in this $M$-cell active BD-RIS, there are $2M$ elements connected to a $2M$-port active reconfigurable impedance network. Specifically, the $2M$-port active reconfigurable impedance network is constructed by a $4M$-port passive reconfigurable impedance network characterized by $\mathbf{\Phi}\in\mathbb{C}^{4M\times 4M}$ and $2M$ amplifiers characterized by $\mathbf{A} = \mathsf{diag}(A_1,\ldots,A_{2M})$, where $A_m> 0$ denotes the power amplification factor of the $m$th reflection-type amplifier, as illustrated in Fig. \ref{fig:RIS_NA}.
	In particular, $2M$ antenna elements are connected to the ports \(1\)-\(2M\) of the \(4M\)-port passive impedance network while $2M$ amplifiers are connected to ports \((2M+1)\)-\(4M\).
	We assume that the $4M$ ports are perfectly matched and isolated among ports \(1\)-\(2M\) and among ports \((2M+1)\)-\(4M\), that is $[\mathbf{\Phi}]_{1:2M;1:2M}=\mathbf{0}$ and $[\mathbf{\Phi}]_{2M+1:4M;2M+1:4M}=\mathbf{0}$ \cite{shen2026active}. This allows us to define $\mathbf{\Theta}\in\mathbb{C}^{2M\times 2M}$ as the scattering matrix of the $2M$-port active reconfigurable impedance network that can be expressed as  
	\begin{equation}\label{eq:unique}
		\mathbf{\Theta} = \mathbf{\Phi}_\mathrm{IA}\mathbf{A}\mathbf{\Phi}_\mathrm{AI},
	\end{equation}
	where $\mathbf{\Phi}_\mathrm{IA} = [\mathbf{\Phi}]_{1:2M;2M+1:4M}\in\mathbb{C}^{2M\times 2M}$ and $\mathbf{\Phi}_\mathrm{AI} = [\mathbf{\Phi}]_{2M+1:4M;1:2M}\in\mathbb{C}^{2M\times 2M}$ are sub-matrices of $\mathbf{\Phi}$ and are unitary when the passive reconfigurable network is lossless. 
    
	According to \cite{shen2026active}, assuming that the multiple antennas at the transmitter/receiver and the elements at the hybrid mode active BD-RIS are perfectly
	matched with no mutual coupling, the corresponding channel matrix between
	the transmitter and each user,  $\mathbf{H}_{k}\in\mathbb{C}^{N_{\mathrm{R},k}\times N_\mathrm{T}}$, $k\in\mathcal{K}$, can be expressed as 
	\begin{equation}
		\mathbf{H}_{k}=\mathbf{H}_{\mathrm{RT},k}+\bar{\mathbf{H}}_{\mathrm{RI},k}\mathbf{\Theta}\bar{\mathbf{H}}_{\mathrm{IT}},\forall k\in\mathcal{K},\label{eq:overall_channel}
	\end{equation}
	where $\mathbf{H}_{\mathrm{RT},k}\in\mathbb{C}^{N_{\mathrm{R},k}\times N_\mathrm{T}}$, $\bar{\mathbf{H}}_{\mathrm{RI},k}\in\mathbb{C}^{N_{\mathrm{R},k}\times 2M}$,
	$\forall k\in\mathcal{K}$, and $\bar{\mathbf{H}}_{\mathrm{IT}}\in\mathbb{C}^{2M\times N_\mathrm{T}}$
	denote channel matrices from the transmitter to user $k$, from active BD-RIS to user $k$, and from the transmitter to active BD-RIS, respectively.
	Then the signal received by user \(k\), $\forall k\in\mathcal{K}$, is 
	\begin{equation}
		\mathbf{y}_{k}=\left(\mathbf{H}_{\mathrm{RT},k}+\bar{\mathbf{H}}_{\mathrm{RI},k}\mathbf{\Theta}\bar{\mathbf{H}}_{\mathrm{IT}}\right)\sum_{k\in\mathcal{K}}\mathbf{F}_{k}\mathbf{s}_{k}+\bar{\mathbf{H}}_{\mathrm{RI},k}\mathbf{\Theta}\mathbf{n}_{\mathrm{I}}+\mathbf{n}_{\mathrm{R},k},
	\end{equation}
	 where $\mathbf{\Theta}\mathbf{n}_{\mathrm{I}}$ is the dynamic noise of active BD-RIS with $\mathbf{n}_\mathrm{I}\sim\mathcal{CN}(\mathbf{0},\sigma_\mathrm{I}^2\mathbf{I}_{2M})$. $\mathbf{s}_{k}\in\mathbb{C}^{N_{\mathrm{S},k}\times 1}$ is the transmit symbol vector for user $k$ with $N_{\mathrm{S},k}$ denoting the number of transmitted symbols, and it satisfies $\mathbb{E}\{\mathbf{s}_k\mathbf{s}_k^\mathsf{H}\} = \mathbf{I}_{N_{\mathrm{S},k}}$, $\forall k\in\mathcal{K}$. $\mathbf{F}_{k}\in\mathbb{C}^{N_\mathrm{T}\times N_{\mathrm{S},k}}$ is the precoder for user $k$ satisfying
	$\sum_{k\in\mathcal{K}}\left\|\mathbf{F}_{k}\right\|_\mathsf{F}^{2}\leq P_{\mathrm{T}}$ with $P_{\mathrm{T}}$ being the transmit power. 
	$\mathbf{n}_{\mathrm{R},k}\sim\mathcal{CN}(\mathbf{0},\sigma_{\mathrm{R},k}^2\mathbf{I}_{N_{\mathrm{R},k}})$ is the noise at user $k$. The constraint of active BD-RIS can be expressed as  
	\begin{equation}
		\| \mathbf{\Theta}\bar{\mathbf{H}}_\mathrm{IT}\mathbf{F}\|_\mathsf{F}^{2}+\sigma_{\mathrm{I}}^{2}\|\mathbf{\Theta}\|_{\mathsf{F}}^{2}\leq P_{\mathrm{A}},\label{eq: active BDRIS constraint}
	\end{equation}
    where $\mathbf{F} = [\mathbf{F}_1,\ldots,\mathbf{F}_K]$ and $P_\mathrm{A}$ denotes the radiated power budget of hybrid mode active BD-RIS.
	
\begin{figure}
		\centering \includegraphics[width=0.48\textwidth]{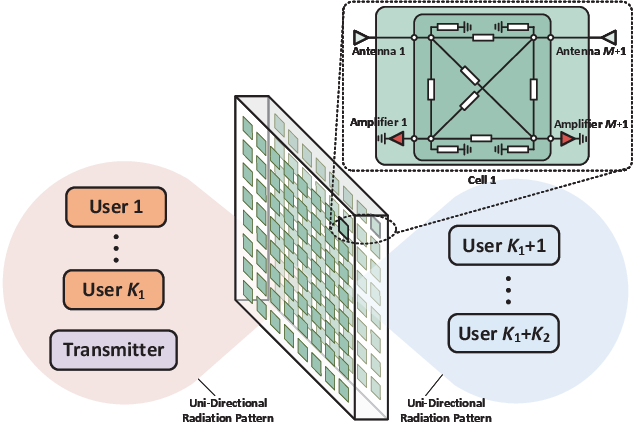}
		\caption{An $M$-cell hybrid mode active BD-RIS with 2$M$ back to back placed uni-directional
			antennas.}
		\label{fig:assumption}
	\end{figure}

	We assume that each element of the active BD-RIS has a uni-directional radiation pattern and that every two elements in one cell are back to back placed to ensure each element covers half of the communication space, as shown in Fig. \ref{fig:assumption}.  
	%     \begin{itemize}
		% 		\item[A1:] Each element has a uni-directional radiation pattern.
		% %		\footnote{In practice the uni-directional radiation pattern can be achieved by using microstrip patch antennas or by adding a reflecting ground plane.}. 
		% 		\item[A2:] In each cell, the two elements are back to back placed so that each
		% 		element covers half space as shown in Fig. \ref{fig:assumption}. 
		% 	\end{itemize}
	Hence, the $M$-cell active BD-RIS has a 2-sector structure, 
	where the first sector consists of elements 1-$M$
	and the other sector consists of elements $(M+1)$-$2M$, and two sectors respectively cover half of the space.
	% The $M$-cell active BD-RIS partitions the whole space into two sides,
	% which are respectively covered by the two sectors. 
	Accordingly, we
	partition the overall channel matrix (\ref{eq:overall_channel}) and
	\(\bar{\mathbf{H}}_{\mathrm{RI},k}\mathbf{\Theta}\mathbf{n}_{\mathrm{I}}\) for user \(k\), \(\forall k\in\mathcal{K}\), as 
	\begin{align}\mathbf{H}_{k}= & \mathbf{H}_{\mathrm{RT},k}+\left[\bar{\mathbf{H}}_{\mathrm{RI},k,1}~\bar{\mathbf{H}}_{\mathrm{RI},k,2}\right]\left[\begin{array}{cc}
			\mathbf{\Theta}_{1,1} & \mathbf{\Theta}_{1,2}\\
			\mathbf{\Theta}_{2,1} & \mathbf{\Theta}_{2,2}
		\end{array}\right]\left[\begin{array}{c}
			\bar{\mathbf{H}}_{\mathrm{IT},1}\\
			\bar{\mathbf{H}}_{\mathrm{IT},2}
		\end{array}\right]\nonumber\\
		= & \mathbf{H}_{\mathrm{RT},k}+\bar{\mathbf{H}}_{\mathrm{RI},k,1}\mathbf{\Theta}_{1,1}\bar{\mathbf{H}}_{\mathrm{IT},1}+\bar{\mathbf{H}}_{\mathrm{RI},k,2}\mathbf{\Theta}_{2,1}\bar{\mathbf{H}}_{\mathrm{IT},1}\nonumber\\
		& +\bar{\mathbf{H}}_{\mathrm{RI},k,1}\mathbf{\Theta}_{1,2}\bar{\mathbf{H}}_{\mathrm{IT},2}+\bar{\mathbf{H}}_{\mathrm{RI},k,2}\mathbf{\Theta}_{2,2}\bar{\mathbf{H}}_{\mathrm{IT},2},\label{eq:overall_channel1}
	\end{align}
	\begin{align}
		\bar{\mathbf{H}}_{\mathrm{RI},k}\mathbf{\Theta}\mathbf{n}_{\mathrm{I}} & =\left[\bar{\mathbf{H}}_{\mathrm{RI},k,1}~\bar{\mathbf{H}}_{\mathrm{RI},k,2}\right]\left[\begin{array}{cc}
			\mathbf{\Theta}_{1,1} & \mathbf{\Theta}_{1,2}\\
			\mathbf{\Theta}_{2,1} & \mathbf{\Theta}_{2,2}
		\end{array}\right]\left[\begin{array}{c}
			\mathbf{n}_{\mathrm{I},1}\\
			\mathbf{n}_{\mathrm{I},2}
		\end{array}\right]\nonumber \\
		& =\bar{\mathbf{H}}_{\mathrm{RI},k,1}\mathbf{\Theta}_{1,1}\mathbf{n}_{\mathrm{I},1}+\bar{\mathbf{H}}_{\mathrm{RI},k,2}\mathbf{\Theta}_{2,1}\mathbf{n}_{\mathrm{I},1}\nonumber \\
		&~~~ +\bar{\mathbf{H}}_{\mathrm{RI},k,1}\mathbf{\Theta}_{1,2}\mathbf{n}_{\mathrm{I},2}+\bar{\mathbf{H}}_{\mathrm{RI},k,2}\mathbf{\Theta}_{2,2}\mathbf{n}_{\mathrm{I},2},\label{eq:dynamic_noise}
	\end{align}
	where $\bar{\mathbf{H}}_{\mathrm{RI},k,i}=[\bar{\mathbf{H}}_{\mathrm{RI},k}]_{:,\mathcal{I}_i}\in\mathbb{C}^{N_{\mathrm{R},k}\times M}$ with $\mathcal{I}_i = (i-1)M+1:iM$ and $\bar{\mathbf{H}}_{\mathrm{IT},i}=[\bar{\mathbf{H}}_\mathrm{IT}]_{\mathcal{I}_i,:}\in\mathbb{C}^{M\times N_\mathrm{T}}$ are channels from sector $i$ of the active BD-RIS
	to user $k$ and from the transmitter to sector $i$ of the
	active BD-RIS, respectively. $\mathbf{n}_{\mathrm{I},i}=[\mathbf{n}_{\mathrm{I}}]_{\mathcal{I}_i}\in\mathbb{C}^{M\times1}$ and $\mathbf{\Theta}_{i,j}=[\mathbf{\Theta}]_{\mathcal{I}_i,\mathcal{I}_j}\in\mathbb{C}^{M\times M}$,
	$\forall i,j\in\{1,2\}$. 
	
	We further assume that the transmitter and $K_1$ users indexed by $\mathcal{K}_{1}=\{1,\ldots,K_{1}\}$, $0<K_{1}<K$, are located within sector 1 of the active BD-RIS,
	and $K_{2}=K-K_{1}$ users indexed by $\mathcal{K}_{2}=\{K_{1}+1,\ldots,K\}$
	are located within sector 2 of the active BD-RIS, as illustrated in Fig. \ref{fig:assumption}. 
	%    To facilitate understanding, we illustrate the locations of the
	% transmitter, active BD-RIS, and users from a top view as shown in
	% Fig. \ref{fig:assumption_TV}. 
	Therefore, the channel from  the transmitter to sector 2 of the active BD-RIS
	is assumed to be zero, that is $\bar{\mathbf{H}}_{\mathrm{IT},2}=\mathbf{0}$. In addition, the channel from one sector of the active BD-RIS to the user from another sector is assumed to be zero, that is $\bar{\mathbf{H}}_{\mathrm{RI},k,i}=\mathbf{0}$,
	$\forall k\in\mathcal{K}_{j}$, $\forall i\ne j$.
	%    Following assumptions A1 and A2, we
	% can deduce the following two corollaries: 
	% \begin{itemize}
		% 	\item[C1:] The channel from the transmitter to sector 2 of the active BD-RIS
		% 	is zero, that is $\bar{\mathbf{H}}_{\mathrm{IT},2}=\mathbf{0}$, because
		% 	the transmitter is not covered by the uni-directional radiation pattern
		% 	of sector 2 of the active BD-RIS. 
		% 	\item[C2:] The channel from sector $i$ of the active BD-RIS to the user belonging
		% 	to $\mathcal{K}_{j}$ is zero $\forall i\ne j$, that is $\bar{\mathbf{H}}_{\mathrm{RI},k,i}=\mathbf{0}$,
		% 	$\forall k\in\mathcal{K}_{j}$, $\forall i\ne j$, because the user
		% 	belonging to $\mathcal{K}_{j}$ is not covered by the uni-directional
		% 	radiation pattern of sector $i$ of the active BD-RIS. 
		% \end{itemize}
	This allows us to simplify the channel matrix
	(\ref{eq:overall_channel1}) and the dynamic noise (\ref{eq:dynamic_noise}) as 
	% \begin{equation}
		% \mathbf{H}_{k}=\left\{ \begin{array}{cc}
			% \mathbf{H}_{\mathrm{RT},k}+\bar{\mathbf{H}}_{\mathrm{RI},k,1}\mathbf{\Theta}_{1,1}\bar{\mathbf{H}}_{\mathrm{IT},1}, & k\in\mathcal{K}_{1},\\
			% \mathbf{H}_{\mathrm{RT},k}+\bar{\mathbf{H}}_{\mathrm{RI},k,2}\mathbf{\Theta}_{2,1}\bar{\mathbf{H}}_{\mathrm{IT},1}, & k\in\mathcal{K}_{2}.
			% \end{array}\right.\label{eq:simplified_overall_channel}
		% \end{equation}
	% \begin{equation}
		% \bar{\mathbf{H}}_{\mathrm{RI},k}\mathbf{\Theta}\mathbf{n}_{\mathrm{I}}=\left\{ \begin{array}{cc}
			% \bar{\mathbf{H}}_{\mathrm{RI},k,1}\mathbf{\Theta}_{1,1}\mathbf{n}_{\mathrm{I},1}\\~~+\bar{\mathbf{H}}_{\mathrm{RI},k,1}\mathbf{\Theta}_{1,2}\mathbf{n}_{\mathrm{I},2}, & k\in\mathcal{K}_{1},\\
			% \bar{\mathbf{H}}_{\mathrm{RI},k,2}\mathbf{\Theta}_{2,1}\mathbf{n}_{\mathrm{I},1}\\~~+\bar{\mathbf{H}}_{\mathrm{RI},k,2}\mathbf{\Theta}_{2,2}\mathbf{n}_{\mathrm{I},2}, & k\in\mathcal{K}_{2}.
			% \end{array}\right.
		% \end{equation}
	\begin{equation}
		\mathbf{H}_{k}=
		\mathbf{H}_{\mathrm{RT},k}+\bar{\mathbf{H}}_{\mathrm{RI},k,i}\mathbf{\Theta}_{i,1}\bar{\mathbf{H}}_{\mathrm{IT},1}, \forall k\in\mathcal{K}_{i},
		\label{eq:simplified_overall_channel}
	\end{equation}
	\begin{equation}
		\bar{\mathbf{H}}_{\mathrm{RI},k}\mathbf{\Theta}\mathbf{n}_{\mathrm{I}}=
		\bar{\mathbf{H}}_{\mathrm{RI},k,i}\mathbf{\Theta}_{i,1}\mathbf{n}_{\mathrm{I},1}+\bar{\mathbf{H}}_{\mathrm{RI},k,i}\mathbf{\Theta}_{i,2}\mathbf{n}_{\mathrm{I},2}, \forall k\in\mathcal{K}_{i}.
	\end{equation}
	Therefore, we can simplify (\ref{eq:received_signal}) as 
	\begin{equation}
		\begin{aligned}
			\mathbf{y}_{k}=&
			(\mathbf{H}_{\mathrm{RT},k}+\bar{\mathbf{H}}_{\mathrm{RI},k,i}\mathbf{\Theta}_{i,1}\bar{\mathbf{H}}_{\mathrm{IT},1})\sum_{k\in\mathcal{K}}\mathbf{F}_{k}\mathbf{s}_{k}\\
			&+\bar{\mathbf{H}}_{\mathrm{RI},k,i}(\mathbf{\Theta}_{i,1}\mathbf{n}_{\mathrm{I},1}+\mathbf{\Theta}_{i,2}\mathbf{n}_{\mathrm{I},2})+\mathbf{n}_{\mathrm{R},k}, \forall k\in\mathcal{K}_{i}.
		\end{aligned}\label{eq:received_signal}
	\end{equation}
	Similarly, we can also partition \eqref{eq: active BDRIS constraint} as 
	\begin{align}
		& \sum_{k\in\mathcal{K}}\| \mathbf{\Theta}\bar{\mathbf{H}}_\mathrm{IT}\mathbf{F}_{k}\|_\mathsf{F}^{2}+\sigma_{\mathrm{I}}^{2}\| \mathbf{\Theta}\|_{\mathsf{F}}^{2}\nonumber \\
		= & \sum_{k\in\mathcal{K}}\left\Vert \left[\begin{array}{cc}
			\mathbf{\Theta}_{1,1} & \mathbf{\Theta}_{1,2}\\
			\mathbf{\Theta}_{2,1} & \mathbf{\Theta}_{2,2}
		\end{array}\right]\left[\begin{array}{c}
			\bar{\mathbf{H}}_{\mathrm{IT},1}\\
			\bar{\mathbf{H}}_{\mathrm{IT},2}
		\end{array}\right]\mathbf{F}_{k}\right\Vert _\mathsf{F}^{2}\non\\
		&~~~+\sigma_{\mathrm{I}}^{2}\left\Vert\left[\begin{array}{cc}
			\mathbf{\Theta}_{1,1} & \mathbf{\Theta}_{1,2}\\
			\mathbf{\Theta}_{2,1} & \mathbf{\Theta}_{2,2}
		\end{array}\right]\right\Vert _{\mathsf{F}}^{2}\nonumber\\
		= & \sum_{k\in\mathcal{K}}\left\Vert \left[\begin{array}{c}
			\mathbf{\Theta}_{1,1}\bar{\mathbf{H}}_{\mathrm{IT},1}\mathbf{F}_{k}+\mathbf{\Theta}_{1,2}\bar{\mathbf{H}}_{\mathrm{IT},2}\mathbf{F}_{k}\\
			\mathbf{\Theta}_{2,1}\bar{\mathbf{H}}_{\mathrm{IT},1}\mathbf{F}_{k}+\mathbf{\Theta}_{2,2}\bar{\mathbf{H}}_{\mathrm{IT},2}\mathbf{F}_{k}
		\end{array}\right]\right\Vert ^{2}_\mathsf{F}\non\\
		&~~~+\sigma_{\mathrm{I}}^{2}\sum_{i}\sum_{j}\left\Vert \mathbf{\Theta}_{i,j}\right\Vert _{\mathsf{F}}^{2}.\label{eq:active_constraint_partioned}
	\end{align}
	
    Using the above assumptions, we can simplify \eqref{eq: active BDRIS constraint} as
	\begin{align}
		&\sum_{k\in\mathcal{K}}\left(\left\Vert \mathbf{\Theta}_{1,1}\bar{\mathbf{H}}_{\mathrm{IT},1}\mathbf{F}_{k}\right\Vert _\mathsf{F}^{2}+\left\Vert \mathbf{\Theta}_{2,1}\bar{\mathbf{H}}_{\mathrm{IT},1}\mathbf{F}_{k}\right\Vert _\mathsf{F}^{2}\right)\nonumber \\
		& \;\;+\sigma_{\mathrm{I}}^{2}\sum_{i}\sum_{j}\left\Vert \mathbf{\Theta}_{i,j}\right\Vert _{\mathsf{F}}^{2}\leq P_{\mathrm{A}}.\label{eq: simplified active constraint}
	\end{align}

	\section{Architecture Design}
	\label{sec:architecture}

	The characteristics of the scattering matrix $\mathbf{\Theta}$ for the $2M$-port active reconfigurable impedance network are determined by the circuit topology of the $4M$-port passive reconfigurable impedance network. When the $4M$-port passive reconfigurable impedance network is reciprocal, we have $\mathbf{\Phi} = \mathbf{\Phi}^\mathsf{T}$, indicating that $\mathbf{\Phi}_\mathrm{IA}=  \mathbf{\Phi}_\mathrm{AI}^\mathsf{T}$ and $\mathbf{\mathbf{\Theta}} = \mathbf{\Theta}^\mathsf{T}$ according to (\ref{eq:unique}). When the $4M$-port passive reconfigurable impedance network is non-reciprocal, its scattering matrix $\mathbf{\Phi}$ does not need to be symmetric, such that $\mathbf{\Phi}_\mathrm{AI}$ and $\mathbf{\Phi}_\mathrm{IA}$ are two independent unitary matrices, resulting in $\mathbf{\Theta}$ an arbitrary complex matrix. The reciprocity will affect the communication model derived in Section \ref{sec:model}, as explained below. 
	
	\subsection{Reciprocal and Non-Reciprocal Active BD-RIS}
	
	\subsubsection{Reciprocal Active BD-RIS}
	We first consider reciprocal active BD-RIS such that $\mathbf{\Theta}_{1,1}=\mathbf{\Theta}_{1,1}^{\mathsf{T}}$,
	$\mathbf{\Theta}_{2,2}=\mathbf{\Theta}_{2,2}^{\mathsf{T}}$, and $\mathbf{\Theta}_{1,2}=\mathbf{\Theta}_{2,1}^{\mathsf{T}}$.
    Observing (\ref{eq:received_signal}), we find that $\mathbf{\Theta}_{2,2}$ will appear only in the denominator of the signal-to-interference-plus-noise ratio (SINR) expression and contribute to the noise term. Therefore, given any feasible $\mathbf{\Theta}_{i,j}$ satisfying (\ref{eq: simplified active constraint}), the optimal $\mathbf{\Theta}_{2,2}$ that maximizes the SINR is given by $\mathbf{\Theta}_{2,2}^\star = \mathbf{0}$.
	
	Furthermore, we denote users covered by sector 1 of the
	active BD-RIS as reflecting users, indexed by an auxiliary notation $\mathcal{K}_{\mathrm{r}}=\mathcal{K}_{1}$ with
	$K_{\mathrm{r}}=K_{1}$. Similarly, we denote users covered by sector 2 as transmitting users, indexed by another auxiliary notation $\mathcal{K}_{\mathrm{t}}=\mathcal{K}_{2}$ with
	$K_{\mathrm{t}}=K_{2}=K-K_{\mathrm{r}}$. Then, using notations
	$\mathbf{\Theta}_{\mathrm{r}}=\mathbf{\Theta}_{1,1}$,
	$\mathbf{\Theta}_{\mathrm{t}}=\mathbf{\Theta}_{2,1}$, 
	$\mathbf{H}_{\mathrm{RI},k}=\bar{\mathbf{H}}_{\mathrm{RI},k,i}$, $\forall k\in\mathcal{K}_{i}$,
	$\forall i\in\{1,2\}$,  $\mathbf{H}_\mathrm{IT}=\bar{\mathbf{H}}_{\mathrm{IT},1}$,
	$\mathbf{n}_{\mathrm{I},1}=\mathbf{n}_{\mathrm{r}}\sim\mathcal{CN}(\mathbf{0},\sigma_\mathrm{I}^2\mathbf{I}_M)$, and $\mathbf{n}_{\mathrm{I},2}=\mathbf{n}_{\mathrm{t}}\sim\mathcal{CN}(\mathbf{0},\sigma_\mathrm{I}^2\mathbf{I}_M)$,
	we can rewrite \eqref{eq:received_signal} and \eqref{eq: simplified active constraint}
	as
	\begin{equation}
		\mathbf{y}_{k}=\left\{ \begin{array}{cc}
			\left(\mathbf{H}_{\mathrm{RT},k}+\mathbf{H}_{\mathrm{RI},k}\mathbf{\Theta}_{\textrm{r}}\mathbf{H}_\mathrm{IT}\right)\sum_{k\in\mathcal{K}}\mathbf{F}_{k}\mathbf{s}_{k}\\
			+\mathbf{H}_{\mathrm{RI},k}\mathbf{\Theta}_{\mathrm{r}}\mathbf{n}_{\mathrm{r}}+\mathbf{H}_{\mathrm{RI},k}\mathbf{\Theta}_{\mathrm{t}}^{\mathsf{T}}\mathbf{n}_{\mathrm{t}}+\mathbf{n}_{\mathrm{R},k}, & k\in\mathcal{K}_{\mathrm{r}},\\
			\left(\mathbf{H}_{\mathrm{RT},k}+\mathbf{H}_{\mathrm{RI},k}\mathbf{\Theta}_{\textrm{t}}\mathbf{H}_\mathrm{IT}\right)\sum_{k\in\mathcal{K}}\mathbf{F}_{k}\mathbf{s}_{k}\\
			+\mathbf{H}_{\mathrm{RI},k}\mathbf{\Theta}_{\mathrm{t}}\mathbf{n}_{\mathrm{r}}+\mathbf{n}_{\mathrm{R},k}, & k\in\mathcal{K}_{\mathrm{t}}.
		\end{array}\right.\label{eq:simplified_overall_channel1}
	\end{equation}
	\begin{align}
		& \left\Vert \mathbf{\Theta}_{\mathrm{r}}\mathbf{H}_\mathrm{IT}\mathbf{F}\right\Vert _\mathsf{F} ^{2}+\left\Vert \mathbf{\Theta}_{\mathrm{t}}\mathbf{H}_\mathrm{IT}\mathbf{F}\right\Vert _\mathsf{F}^{2}\nonumber \\
		& \;\;\;\;+\sigma_{\mathrm{I}}^{2}\left(\left\Vert \mathbf{\Theta}_{\mathrm{r}}\right\Vert _{\mathsf{F}}^{2}+2\left\Vert \mathbf{\Theta}_{\mathrm{t}}\right\Vert _{\mathsf{F}}^{2}\right)\leq P_{\mathrm{A}},\label{eq: simplified active constraint-1}
	\end{align}
	where $\mathbf{\Theta}_{\mathrm{r}}=\mathbf{\Theta}_{\mathrm{r}}^{\mathsf{T}}$
	can be any complex symmetric matrices and $\mathbf{\Theta}_{\mathrm{t}}$
	can be any matrices. 
	
	% \begin{figure}
		% 	\centering \includegraphics[height=2.1 in]{figures/assumptions_TV}
		% 	\caption{Top view for the locations of the $M$-cell BD-RIS partitioning the
			% 		whole space into two sides, the transmitter, and multiple users.}
		% 	\label{fig:assumption_TV}
		% \end{figure}

	\subsubsection{Non-Reciprocal Active BD-RIS}
	We next consider non-reciprocal active BD-RIS such that $\mathbf{\Theta}_{1,1}$,
	$\mathbf{\Theta}_{1,2}$, $\mathbf{\Theta}_{2,1}$, and $\mathbf{\Theta}_{2,2}$
	can be any matrices. 
	We find from (\ref{eq:received_signal}) that $\mathbf{\Theta}_{1,2}$ and $\mathbf{\Theta}_{2,2}$ appear only in the denominator of the SINR expression and contribute to the noise term. Therefore, given any feasible $\mathbf{\Theta}_{i,j}$ satisfying (\ref{eq: simplified active constraint}), the optimal $\mathbf{\Theta}_{1,2}$ and $\mathbf{\Theta}_{2,2}$ that maximize SINR are respectively given by $\mathbf{\Theta}_{1,2}^\star = \mathbf{0}$ and $\mathbf{\Theta}_{2,2}^\star = \mathbf{0}$.
	
	Similarly, we can rewrite \eqref{eq:received_signal}
	and \eqref{eq: simplified active constraint} as
	% \begin{equation}
	% 	\mathbf{y}_{k}=\left\{ \begin{array}{cc}
	% 		\left(\mathbf{H}_{\mathrm{RT},k}+\mathbf{H}_{\mathrm{RI},k}\mathbf{\Theta}_{\mathrm{r}}\mathbf{H}_\mathrm{IT}\right)\sum_{k\in\mathcal{K}}\mathbf{F}_{k}\mathbf{s}_{k}\\
	% 		+\mathbf{H}_{\mathrm{RI},k}\mathbf{\Theta}_{\mathrm{r}}\mathbf{n}_{\mathrm{r}}+\mathbf{n}_{\mathrm{R},k}, & k\in\mathcal{K}_{\mathrm{r}},\\
	% 		\left(\mathbf{H}_{\mathrm{RT},k}+\mathbf{H}_{\mathrm{RI},k}\mathbf{\Theta}_{\mathrm{t}}\mathbf{H}_\mathrm{IT}\right)\sum_{k\in\mathcal{K}}\mathbf{F}_{k}\mathbf{s}_{k}\\
	% 		+\mathbf{H}_{\mathrm{RI},k}\mathbf{\Theta}_{\mathrm{t}}\mathbf{n}_{\mathrm{r}}+\mathbf{n}_{\mathrm{R},k}, & k\in\mathcal{K}_{\mathrm{t}}.
	% 	\end{array}\right.\label{eq:simplified_overall_channel1-1}
	% \end{equation}
    \begin{equation}
        \begin{aligned}
		\mathbf{y}_{k}=		&\left(\mathbf{H}_{\mathrm{RT},k}+\mathbf{H}_{\mathrm{RI},k}\mathbf{\Theta}_{w}\mathbf{H}_\mathrm{IT}\right)\sum_{k\in\mathcal{K}}\mathbf{F}_{k}\mathbf{s}_{k}\\
			&+\mathbf{H}_{\mathrm{RI},k}\mathbf{\Theta}_{w}\mathbf{n}_{\mathrm{r}}+\mathbf{n}_{\mathrm{R},k}, \forall k\in\mathcal{K}_{w},\forall w\in\{\mathrm{r,t}\},
        \end{aligned}\label{eq:simplified_overall_channel1-1}
	\end{equation}
	\begin{align}
		& \left\Vert \mathbf{\Theta}_{\mathrm{r}}\mathbf{H}_\mathrm{IT}\mathbf{F}\right\Vert _\mathsf{F}^{2}+\left\Vert \mathbf{\Theta}_{\mathrm{t}}\mathbf{H}_\mathrm{IT}\mathbf{F}\right\Vert _\mathsf{F}^{2}\nonumber \\
		& \;\;\;\;+\sigma_{\mathrm{I}}^{2}\left(\left\Vert \mathbf{\Theta}_{\mathrm{r}}\right\Vert _{\mathsf{F}}^{2}+\left\Vert \mathbf{\Theta}_{\mathrm{t}}\right\Vert _{\mathsf{F}}^{2}\right)\leq P_{\mathrm{A}},\label{eq: simplified active constraint-1-1}
	\end{align}
	where $\mathbf{\Theta}_{\mathrm{r}}$ and $\mathbf{\Theta}_{\mathrm{t}}$
	can be any matrices. 
	
	\subsection{Cell-Wise Architectures}

	\begin{figure*}[!t]
		\centering
		\subfigure[A 2-cell reciprocal active BD-RIS with cell-wise single-connected architecture]{
			\includegraphics[width=0.3\textwidth]{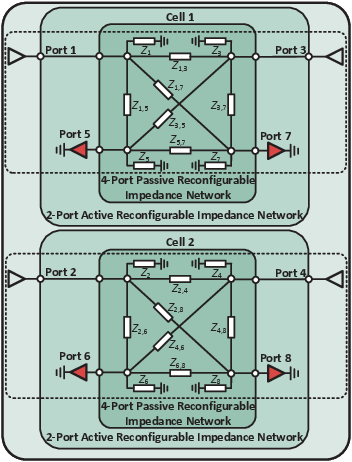}\label{fig:single}}
		\subfigure[A 4-cell active BD-RIS with cell-wise fully-connected architecture]{
			\includegraphics[width=0.3\textwidth]{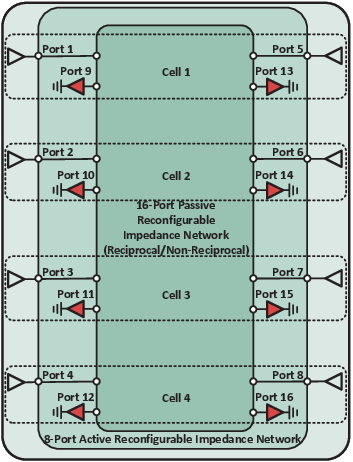}\label{fig:fully}}
		\subfigure[A 4-cell active BD-RIS with cell-wise group-connected architecture (\(G=2\))]{
			\includegraphics[width=0.3\textwidth]{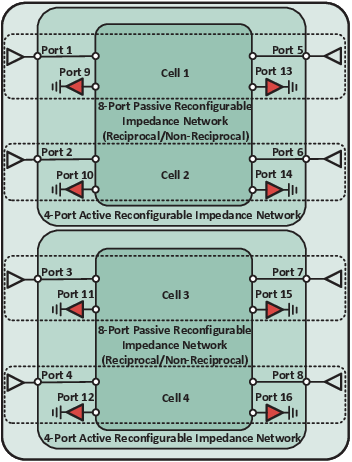}\label{fig:group}}
		\caption{Different cell-wise architectures of active BD-RIS with transmitting and reflecting mode.}
	\end{figure*}

	% The circuit topology of the $4M$-port passive reconfigurable impedance network determines the characteristic of the scattering matrices $\mathbf{\Theta}_\mathrm{r}$ and $\mathbf{\Theta}_\mathrm{t}$ of active BD-RIS with hybrid mode, which can  be explicitly written as
	% \begin{equation}\label{eq:unique3}
	% 	\mathbf{\Theta}_{\mathrm{r}}=\mathbf{\Phi}_{\mathrm{IA},11}\mathbf{A}_1\mathbf{\Phi}_{\mathrm{AI},11}+\mathbf{\Phi}_{\mathrm{IA},12}\mathbf{A}_2\mathbf{\Phi}_{\mathrm{AI},21},
	% \end{equation}
	% \begin{equation}\label{eq:unique4}
	% 	\mathbf{\Theta}_{\mathrm{t}}=\mathbf{\Phi}_{\mathrm{IA},21}\mathbf{A}_1\mathbf{\Phi}_{\mathrm{AI},11}+\mathbf{\Phi}_{\mathrm{IA},22}\mathbf{A}_2\mathbf{\Phi}_{\mathrm{AI},21},
	% \end{equation}    
	% where $\mathbf{\Phi}_{\mathrm{IA},ij}=[\mathbf{\Phi}_\mathrm{IA}]_{\mathcal{I}_i,\mathcal{I}_j}\in\mathbb{C}^{M\times M}$, $\mathbf{A}_{i}=[\mathbf{A}]_{\mathcal{I}_i,\mathcal{I}_i}\in\mathbb{C}^{M\times M}$, $\mathbf{\Phi}_{\mathrm{AI},ij}=[\mathbf{\Phi}_\mathrm{AI}]_{\mathcal{I}_i,\mathcal{I}_j}\in\mathbb{C}^{M\times M}$, $\forall i,j\in\{1,2\}$. 
    In this section, we introduce three representative cell-wise architectures of hybrid mode active BD-RIS, namely cell-wise single, fully, and group-connected that are determined by the inter-cell circuit topologies, and derive the corresponding constraints of $\mathbf{\Theta}$. 
	
	\subsubsection{Cell-Wise Single-Connected}
	The cell-wise single-connected active BD-RIS consists of isolated cells, each consisting of a 4-port passive reconfigurable impedance network with two ports being connected to two back to back placed antennas and the rest being connected to two amplifiers. An example of a reciprocal cell-wise single-connected architecture can be found in Fig. \ref{fig:single}.
	In this case, both $\mathbf{\Theta}_\mathrm{r}$ and $\mathbf{\Theta}_\mathrm{t}$ are diagonal matrices, that is 
	\begin{subequations}\label{eq:single}
		\begin{align}		
			\mathbf{\Theta}_\mathrm{r}=\mathsf{diag}\left(\Theta_{\mathrm{r},1},\Theta_{\mathrm{r},2},\ldots,\Theta_{\mathrm{r},M}\right),\\
			\mathbf{\Theta}_\mathrm{t}=\mathsf{diag}\left(\Theta_{\mathrm{t},1},\Theta_{\mathrm{t},2},\ldots,\Theta_{\mathrm{t},M}\right).
		\end{align}
	\end{subequations}  

	\begin{remark}    
    Prior works on active STAR-RISs \cite{xu2023,Maghrebi2024,Faramarzi2025,wang2026,aung2025,rongguang2025,huang2025} and active double-faced RISs \cite{Liu2022c,Guo2023,Zhou2023a} directly introduce amplification factors and phase shifters into the model to obtain (\ref{eq:single}), thus neglecting the intrinsic electromagnetic characteristics introduced by reflection-type amplifiers and the reconfigurable impedance network. 
    In contrast, this work explicitly incorporates these characteristics. Based on equation (\ref{eq:unique}), we write $\Theta_{\mathrm{r},i}$ and $\Theta_{\mathrm{t},i}$, $\forall i\in\mathcal{M}$, as 
    \begin{subequations}\label{eq:unique3}
        \begin{align}
		\Theta_{\mathrm{r},i}=&[\mathbf{\Phi}_\mathrm{IA}]_{i,i}A_i[\mathbf{\Phi}_\mathrm{AI}]_{i,i}\\
        &~~~+[\mathbf{\Phi}_\mathrm{IA}]_{i,M+i}A_{M+i}[\mathbf{\Phi}_\mathrm{AI}]_{M+i,i},\nonumber\\
		\Theta_{\mathrm{t},i}=&[\mathbf{\Phi}_\mathrm{IA}]_{M+i,i}A_i[\mathbf{\Phi}_\mathrm{AI}]_{i,i}\nonumber\\
        &~~~+[\mathbf{\Phi}_\mathrm{IA}]_{M+i,M+i}A_{M+i}[\mathbf{\Phi}_\mathrm{AI}]_{M+i,i}.
        \end{align}
	\end{subequations}  
    This reveals that the magnitude gain of the signal scattered by active BD-RIS toward either the reflecting sector or the transmitting sector is jointly determined by the amplification factors of all reflection-type amplifiers.
	\end{remark}

	\subsubsection{Cell-Wise Fully-Connected}
	In this architecture, $M$ cells are interconnected via a $2M$ port active reconfigurable impedance network that consists of a $4M$ passive reconfigurable impedance network, whose half ports are connected to antenna elements and the rest are connected to reflection-type amplifiers.
	An instance of this architecture is illustrated in Fig. \ref{fig:fully}.
	In this case, $\mathbf{\Theta}$ can be any complex matrices such that both $\mathbf{\Theta}_\mathrm{r}$ and $\mathbf{\Theta}_\mathrm{t}$ are any complex matrices. 
	In particular, when the passive reconfigurable impedance network is reciprocal, we have the constraint as
	\begin{equation}\label{fully_con}
		\mathbf{\Theta}_\mathrm{r}=\mathbf{\Theta}_\mathrm{r}^\mathsf{T}, \mathbf{\Theta}_\mathrm{r}\in\mathbb{C}^{M\times M},\mathbf{\Theta}_\mathrm{t}\in\mathbb{C}^{M\times M},
	\end{equation}
	meaning that \(\mathbf{\Theta}_\mathrm{r}\) can be any complex symmetric matrices, while \(\mathbf{\Theta}_\mathrm{t}\) can be any complex matrices. When the passive reconfigurable impedance network is non-reciprocal, constraint \eqref{fully_con} is released and $\mathbf{\Theta}_\mathrm{r}$ and $\mathbf{\Theta}_\mathrm{t}$ can be any complex matrices.
	
	\subsubsection{Cell-Wise Group-Connected}
	In this architecture, $M$ cells of the hybrid mode active BD-RIS are uniformly partitioned into $\bar{G}$ interlaced groups indexed by \(g \in \mathcal{G}\triangleq \left\{1,\ldots,\bar{G}\right\}\). Therefore, each group contains $\bar{G} = \frac{M}{\bar{G}}$ cells that form a cell-wise fully-connected architecture. One example of such an architecture can be found in Fig. \ref{fig:group}. 
	In this case, both $\mathbf{\Theta}_\mathrm{r}$ and $\mathbf{\Theta}_\mathrm{t}$ are block-diagonal matrices. 
	For reciprocal architectures, \(\mathbf{\Theta}_\mathrm{r}\) and \(\mathbf{\Theta}_\mathrm{t}\) are subject to the following constraints
	\begin{equation}
		\begin{aligned}
			\mathbf{\Theta}_\mathrm{r}&=\mathsf{blkdiag}\left(\mathbf{\Theta}_\mathrm{r,1},\ldots,\mathbf{\Theta}_{\mathrm{r},\bar{G}}\right), \mathbf{\Theta}_{\mathrm{r},g}=\mathbf{\Theta}_{\mathrm{r},g}^\mathsf{T}, \forall g \in \mathcal{G},\\
			\mathbf{\Theta}_\mathrm{t}&=\mathsf{blkdiag}\left(\mathbf{\Theta}_\mathrm{t,1},\ldots,\mathbf{\Theta}_{\mathrm{t},\bar{G}}\right), \forall g \in \mathcal{G},
		\end{aligned}
	\end{equation}
	where \(\mathbf{\Theta}_{\mathrm{r},g} \in \mathbb{C}^{G\times G}\) and \(\mathbf{\Theta}_{\mathrm{t},g}\) represent the scattering matrices of the \(g\)th group. 
	For non-reciprocal architectures, the symmetric constraints of $\mathbf{\Theta}_{\mathrm{r},g}$, $\forall g\in\mathcal{G}$, are released such that both $\mathbf{\Theta}_\mathrm{r}$ and $\mathbf{\Theta}_\mathrm{t}$ are block diagonal with each block being any complex matrices.
	The cell-wise group-connected architecture is more generalized since it can become cell-wise single-connected when \(G=1\) and become cell-wise fully-connected when \(G=M\). 

    The proposed architectures and corresponding constraints of $\mathbf{\Theta}_\mathrm{r}$ and $\mathbf{\Theta}_\mathrm{t}$ are summarized in Table \ref{tab:architecture}. 

      \begin{table*}[]
		\centering
            \caption{Architectures and Constraints of Hybrid Mode Active BD-RIS}\label{tab:architecture}
		\begin{tabular}{|c|c|c|c|}
			\hline
			     Architectures & Cell-Wise Single-Connected & Cell-Wise Fully-Connected & Cell-Wise Group-Connected  \\ 
            \hline
			Reciprocal  &   \multirow{4}{*}{\begin{tabular}[c]{@{}c@{}} $\mathbf{\Theta}_\mathrm{r}=\mathsf{diag}\left(\Theta_{\mathrm{r},1},\Theta_{\mathrm{r},2},\ldots,\Theta_{\mathrm{r},M}\right)$ \\ $\mathbf{\Theta}_\mathrm{t}=\mathsf{diag}\left(\Theta_{\mathrm{t},1},\Theta_{\mathrm{t},2},\ldots,\Theta_{\mathrm{t},M}\right)$ \end{tabular}}   &  \begin{tabular}[c]{@{}c@{}} $\mathbf{\Theta}_\mathrm{r}=\mathbf{\Theta}_\mathrm{r}^\mathsf{T}, \mathbf{\Theta}_\mathrm{r}\in\mathbb{C}^{M\times M}$\\ $\mathbf{\Theta}_\mathrm{t}\in\mathbb{C}^{M\times M}$\end{tabular}  & \begin{tabular}[c]{@{}c@{}} $\mathbf{\Theta}_\mathrm{r}=\mathsf{blkdiag}\left(\mathbf{\Theta}_\mathrm{r,1},\ldots,\mathbf{\Theta}_{\mathrm{r},\bar{G}}\right), \mathbf{\Theta}_{\mathrm{r},g}=\mathbf{\Theta}_{\mathrm{r},g}^\mathsf{T}$\\ $\mathbf{\Theta}_\mathrm{t}=\mathsf{blkdiag}\left(\mathbf{\Theta}_\mathrm{t,1},\ldots,\mathbf{\Theta}_{\mathrm{t},\bar{G}}\right)$ \end{tabular}\\
            \cline{1-1}\cline{3-4}
            Non-Reciprocal & & $\mathbf{\Theta}_\mathrm{r}\in\mathbb{C}^{M\times M},~\mathbf{\Theta}_\mathrm{t}\in\mathbb{C}^{M\times M}$ & \begin{tabular}[c]{@{}c@{}} $\mathbf{\Theta}_\mathrm{r}=\mathsf{blkdiag}\left(\mathbf{\Theta}_\mathrm{r,1},\ldots,\mathbf{\Theta}_{\mathrm{r},\bar{G}}\right)$\\ $\mathbf{\Theta}_\mathrm{t}=\mathsf{blkdiag}\left(\mathbf{\Theta}_\mathrm{t,1},\ldots,\mathbf{\Theta}_{\mathrm{t},\bar{G}}\right)$\end{tabular}\\
            \hline
		\end{tabular}	
	\end{table*}

	\begin{remark}
		Active STAR-RISs \cite{xu2023,Maghrebi2024,Faramarzi2025,wang2026,aung2025,rongguang2025,huang2025} and active double-faced RISs \cite{Liu2022c,Guo2023,Zhou2023a} are essentially special cases of the reciprocal hybrid mode active BD-RIS with cell-wise single-connected architecture, as illustrated in Fig. \ref{fig:single}. 
    \end{remark}
    \begin{remark}
        When no inter-connections exist between elements within each cell and all antennas are placed in the same direction, the proposed architecture becomes the active D-RIS \cite{prevail2023,partialCSI2024,allu2023robust,Yaswanth2024,Chen2022a,Ye2025,Zhu2023a,Yue2024,Peng2024,Lyu2023,Zhai2024,ming2024}.
	\end{remark}

	\section{Joint Precoding and Active BD-RIS Design}
	\label{sec:joint optimization}
	
	In this section, we formulate the sum rate maximization problem for reciprocal and non-reciprocal hybrid mode active BD-RIS aided multiuser systems based on the above model and further propose a unified optimization framework that is applicable to all the architectures proposed in Section \ref{sec:architecture}.
	
	\subsection{Problem Formulation}\label{active}
	We consider a MU-MISO communication system assisted by a hybrid mode active BD-RIS.
	The system comprises of an $N_\mathrm{T}$-antenna base station (BS), an $M$-cell active BD-RIS and $K=K_\mathrm{r}+K_\mathrm{t}$ single-antenna users randomly located in both sides of the hybrid mode active BD-RIS based on Section \ref{sec:model}. Then the relevant notations in Section \ref{sec:architecture} become $\mathbf{H}_{\mathrm{RT},k} = \mathbf{h}_{\mathrm{RT},k}\in\mathbb{C}^{1\times N_\mathrm{T}}$, $\mathbf{H}_{\mathrm{RI},k} = \mathbf{h}_{\mathrm{RI},k}\in\mathbb{C}^{1\times M}$, $\forall k\in\mathcal{K}$, $\mathbf{F} = [\mathbf{f}_1,\ldots,\mathbf{f}_K]\in\mathbb{C}^{N_\mathrm{T}\times K}$,  $\mathbf{s}_k = s_k\in\mathbb{C}$, and $\mathbf{n}_{\mathrm{R},k} = n_{\mathrm{R},k}\in\mathbb{C}$, $\forall k\in\mathcal{K}$. Under the assumption that the BS perfectly knows channel state information (CSI), we aim to explore the performance boundary of the proposed hybrid mode active BD-RIS aided MU-MISO system.
	
	Define \( \mathbf{h} _{w,k}\triangleq \mathbf{h}_{\mathrm{RT},k}+\mathbf{h}_{\mathrm{RI},k} \mathbf{\Theta} _{w}\mathbf{H}_{\mathrm{IT}} \),  \(\forall w \in \left\{\mathrm{t}, \mathrm{r}\right\}\), \( \forall k \in \mathcal{K}_w \). The SINR for a given user in the presence of an active BD-RIS is calculated as follows
	\begin{equation}
		\gamma_{k}=\frac{A_{w,k}}{B_{w,k}}, \qquad \forall w \in \left\{\mathrm{t}, \mathrm{r}\right\},  \forall k \in \mathcal{K}_w, 
	\end{equation}
	where we define 
	\begin{equation}
		A_{w,k} \triangleq|\mathbf{h}_{w,k}\mathbf{f}_k|^2,~~ \forall w\in\{\mathrm{r,t}\}, \forall k\in\mathcal{K}_w,
	\end{equation}
	in the presence of either reciprocal or non-reciprocal active BD-RIS. Meanwhile, we define
	\begin{align}
		&	B_{\mathrm{r},k}\triangleq\notag\\
		&\left\{ \begin{array}{ll}
			\sum_{\substack{i\in\mathcal{K}\\i\ne k }}\left|\mathbf{h} _{\mathrm{r},k}\mathbf{f}_{i}\right|^2
			+ \sigma_{\mathrm{I}}^{2}\left\|\mathbf{h}_{\mathrm{RI},k} \mathbf{\Theta} _{\mathrm{r}}\right\|_2^2 &{}\\
			~~~~~+ \sigma_{\mathrm{I}}^{2}\left\|\mathbf{h}_{\mathrm{RI},k} \mathbf{\Theta} _{\mathrm{t}}^{\mathsf{T}}\right\|_2^2
			+ \sigma_{\mathrm{R},k}^{2}, & \text{reciprocal},\\
			\sum\limits_{\substack{i\in\mathcal{K}\\i\ne k }}\left|\mathbf{h}_{\mathrm{r},k}\mathbf{f}_{i}\right|_2^2
			+ \sigma_{\mathrm{I}}^{2}\left\|\mathbf{h}_{\mathrm{RI},k} \mathbf{\Theta} _{\mathrm{r}}\right\|_2^2 
			+ \sigma_{\mathrm{R},k}^{2}, &\text{non-reciprocal},
		\end{array}\right.\label{eq:simplified_sinr_1}
	\end{align}
	\begin{equation}
		B_{\mathrm{t},k}\triangleq \sum_{i\in\mathcal{K},i\ne k }\left|\mathbf{h} _{\mathrm{t},k}\mathbf{f}_{i}\right|^2
		+ \sigma_{\mathrm{I}}^{2}\left\|\mathbf{h}_{\mathrm{RI},k} \mathbf{\Theta} _{\mathrm{t}}\right\|^2 
		+ \sigma_{\mathrm{R},k}^{2}.
	\end{equation}

    The optimization problem for sum-rate maximization is formulated as 
	\begin{subequations}\label{eq:objective_1}
		\begin{align}
			\max_{\mathbf{F},\mathbf{\Theta}_{\mathrm{r}},\mathbf{\Theta}_{\mathrm{t}}} &f\left (\mathbf{F},\mathbf{\Theta}_{\mathrm{r}},\mathbf{\Theta}_{\mathrm{t}}  \right ) =
			\sum_{k \in \mathcal{K} }\log_{2}\left ( {1+\gamma_{k}}  \right ) \label{eq: obj_1} \\
			\text{s.t.} \;\;\;\;&\eqref{eq: simplified active constraint-1} \;\text{and}\; \mathbf{\Theta}_{\mathrm{r}}=\mathbf{\Theta}_{\mathrm{r}}^{\mathsf{T}}\; \text{for reciprocal BD-RIS or }\notag \\
			\;\;\;\;\qquad&\eqref{eq: simplified active constraint-1-1}\; \text{for non-reciprocal BD-RIS}\label{eq:objective_1_st2},\\
			&\left\|\mathbf{F}\right\|_\mathsf{F}^{2}\leq P_{\mathrm{T}}.\label{eq:objective_1_st3}
		\end{align}
	\end{subequations}
    Given the intractable objective function \eqref{eq: obj_1} and the constraints, which comprise logarithmic and fractional terms, we will reformulate them into multi-variable/block problems and address each variable or block iteratively in the following subsections.

	\subsection{A Unified Optimization Framework}\label{reci_0}
	We propose a unified optimization framework for reciprocal and non-reciprocal hybrid mode active BD-RIS.
	Based on the \textit{Lagrangian Dual Transformation} \cite{FPI},
	we start by moving the fractional terms \( \gamma_{k} \), \( \forall k \in \mathcal{K}  \) out of the logarithmic function
	and transforming \( f\left (\mathbf{F},\mathbf{\Theta}_{\mathrm{r}},\mathbf{\Theta}_{\mathrm{t}}  \right ) \) into 
	\begin{align}
		f_{\iota}&\left (\mathbf{F},\mathbf{\Theta}_{\mathrm{r}},\mathbf{\Theta}_{\mathrm{t}},\bm{\iota}\right )= \sum_{w \in \left\{\mathrm{t}, \mathrm{r}\right\}}\sum_{k \in \mathcal{K}_w }
		\bigg(  \log_{2}\left ( {1+\iota_{k}}  \right )-\iota_{k}\big.\notag\\
		&\qquad\qquad\qquad\qquad\qquad\qquad\;\;\big.+\frac{\left ( 1+\iota_{k} \right )A_{w,k} }
		{A_{w,k}+B_{w,k}}  \bigg) \label{eq:objective_dual},
	\end{align}
	where \( \bm{\iota} \triangleq [\iota_1,\ldots,\iota_{K} ]^\mathsf{T}\in \mathbb{R}^{K\times 1}\) is an auxiliary vector.
	Observing \eqref{eq:objective_dual}, the fractions are independent of logarithmic function compared with the original objective function \eqref{eq: obj_1}.
	We next apply the \textit{Quadratic Transform} \cite{FPI} to transform fractional terms into integral expressions 
	\begin{align}
		f_{\tau}&\left (\mathbf{F},\mathbf{\Theta}_{\mathrm{r}},\mathbf{\Theta}_{\mathrm{t}},\bm{\iota},\bm{\tau}\right )=
		\sum_{w \in \left\{\mathrm{t}, \mathrm{r}\right\}}\sum_{k \in \mathcal{K}_w }
		\bigg ( \log_{2}\left ( {1+\iota_{k}}  \right )-\iota_{k} \notag\\
		&\qquad+ 2\sqrt{1+\iota_k} \Re\left \{ \tau_k^* \mathbf{h}_{w,k}\mathbf{f}_k \right \}-|\tau_k|^2 \left(A_{w,k}+B_{w,k}\right) \bigg), \label{eq:objective_quadratic}
	\end{align}
	where \( \bm{\tau} \triangleq [\tau_1,\ldots,\tau_{K} ]^\mathsf{T}\in \mathbb{C}^{K\times 1}\) is another auxiliary vector.
	With two new auxiliary vectors, problems \eqref{eq:objective_1} can be reformulated into the following unified form
	\begin{subequations}\label{eq:objective_transformed}
		\begin{align}
			\max_{\mathbf{F},\mathbf{\Theta}_{\mathrm{r}},\mathbf{\Theta}_{\mathrm{t}},\bm{\iota},\bm{\tau}} &f_{\tau}\left (\mathbf{F},\mathbf{\Theta}_{\mathrm{r}},\mathbf{\Theta}_{\mathrm{t}},\bm{\iota},\bm{\tau}  \right ) \\
			\text{s.t.} \;\;\;\;\;\; &\eqref{eq:objective_1_st2}\text{-}\eqref{eq:objective_1_st3}.
		\end{align}
	\end{subequations}

    Due to the multi-block structure, problem \eqref{eq:objective_transformed} can be solved efficiently by resorting to the block coordinate descent (BCD) algorithm \cite{DBertsekas}.
	 The proposed unified joint transmit precoding and hybrid mode active BD-RIS design is summarized in Algorithm \ref{alg:joint_precoder_bdris}. 
	We will elaborate on the solutions for each block in the following subsections. 
	Section \ref{reci_1} introduces the solutions to auxiliary vectors $\bm{\iota}$ and $\bm{\tau}$; Section \ref{reci_2} derives the solution for the transmit precoder $\mathbf{F}$; and Section \ref{reci_3} addresses the design of either reciprocal or non-reciprocal active BD-RIS.
	
	\begin{algorithm}[t]
		\caption{Joint Transmit Precoding and Hybrid mode Active BD-RIS Design}
		\label{alg:joint_precoder_bdris}
		\begin{algorithmic}[1]
			\REQUIRE $\mathbf{h}_{\mathrm{RT},k},\mathbf{h}_{\mathrm{RI},k},
			\sigma_{\mathrm{R},k}, \forall k\in\mathcal{K},\mathbf{H}_{\mathrm{IT}}, \sigma_\mathrm{I}, P_\mathrm{T},P_\mathrm{A}.$
			\ENSURE $\mathbf{\Theta}_\mathrm{r}^\star,\;\mathbf{\Theta}_\mathrm{t}^\star,\ \mathbf{F}^\star.$
			
			\STATE Initialize $\mathbf{\Theta}_\mathrm{r},\mathbf{\Theta}_\mathrm{t}, \mathbf{F}.$
			\WHILE{$\text{objective \eqref{eq: obj_1} not converged}$}
			\STATE Update $\bm{\iota}^\star$ by \eqref{eq:iota}.
			\STATE Update $\bm{\tau}^\star$ by \eqref{eq:tau}.
			\STATE Update $\mathbf{F}^\star$ by \eqref{eq:F}.
			\STATE Update $\mathbf{\Theta}_\mathrm{r/t}^\star$ by reciprocal formulation \eqref{eq:reci_theta_r} and \eqref{eq:reci_theta_t} or by non-reciprocal formulation \eqref{eq:nonreci_theta}.
			\ENDWHILE
			% \RETURN $\mathbf{\Theta}_\mathrm{r}^\star,\;\mathbf{\Theta}_\mathrm{t}^\star,\ \mathbf{F}^\star.$
		\end{algorithmic}
	\end{algorithm}
	\subsubsection{Auxiliary Vectors $\bm{\iota}$ and $\bm{\tau}$ }\label{reci_1}
	When blocks \(\mathbf{F},\mathbf{\Theta}_{\mathrm{r}},\mathbf{\Theta}_{\mathrm{t}}\) are fixed, the sub-problems associated with $\bm{\iota}$ and $\bm{\tau}$ are convex optimization problems with no constraints, which enable us to derive the optimal solution as
	%    the optimal value of \(\bm{\iota}^\star\) (or \(\bm{\tau}^\star\)) directly
	% by setting \(\frac{\partial f_{\tau}\left(\mathbf{W}, \boldsymbol{\Phi}_{\mathrm{t}}, \boldsymbol{\Phi}_{\mathrm{r}}, \boldsymbol{\iota}, \boldsymbol{\tau}\right)}{\partial \boldsymbol{\iota}}=\mathbf{0}\) ( or \(\frac{\partial f_{\tau}\left(\mathbf{W}, \boldsymbol{\Phi}_{\mathrm{t}}, \boldsymbol{\Phi}_{\mathrm{r}}, \boldsymbol{\iota}, \boldsymbol{\tau}\right)}{\partial \boldsymbol{\tau}}=\mathbf{0}\)). We can obtain the solutions to auxiliary blocks for reciprocal and non-reciprocal active BD-RIS simultaneously as follows:
	\begin{equation}\label{eq:iota}
		\iota_k^\star=\gamma_{k}, \;\forall k \in \mathcal{K},
	\end{equation}
	\begin{equation}\label{eq:tau}
		\tau_{k}^\star=\frac{\sqrt{1+\iota_k} \mathbf{h}_{w,k}\mathbf{f}_k }
		{A _{w,k}+B _{w,k}}, \;\;\forall w \in \left\{\mathrm{t},\mathrm{r}\right\}, \;\forall k \in \mathcal{K}_w.
	\end{equation}
	
	\subsubsection{Transmit Precoder $\mathbf{F}$ }\label{reci_2}
	When blocks \(\mathbf{\Theta}_{\mathrm{r}},\mathbf{\Theta}_{\mathrm{t}},\bm{\iota},\bm{\tau}\) are fixed, the sub-problem associated with transmit precoder $\mathbf{F}$ can be simplified as 
	\begin{subequations}\label{eq:Reciprocal_F}
		\begin{align}
			\max_{\mathbf{F} }
			&\sum_{w\in \left\{\mathrm{t},\mathrm{r}\right\}}\sum_{k \in \mathcal{K}_w }\bigg (   2\sqrt{1+\iota_k}  \Re\left \{ \tau_{k}^*\mathbf{h}_{w,k}\mathbf{f}_{k} \right \}\bigg. \notag\\
			&\qquad\qquad\qquad\qquad\;\;\;- \bigg.|\tau_k|^2 \sum\limits_{i\in\mathcal{K} }\left|\mathbf{h} _{w,k}\mathbf{f}_{i}\right|^2\bigg)
			\\
			\text{s.t.} \;\;   &\eqref{eq: simplified active constraint-1} \;\text{and}\; \eqref{eq:objective_1_st3} \;\text{for reciprocal BD-RIS or}\notag\\
			\;\;\;\;\; \; \; \; \;  &\eqref{eq: simplified active constraint-1-1} \;\text{and}\; \eqref{eq:objective_1_st3} \;\text{for non-reciprocal BD-RIS}.
		\end{align} 
	\end{subequations}
	Further, defining $\bar{\mathbf{h}}_k \triangleq \tau_{k}^*\mathbf{h} _{w,k}$, $\forall w\in \left \{ \mathrm{t},\mathrm{r}\right \}$, $ \forall k\in \mathcal{K}_w$ and $\tilde{\mathbf{\Theta}} \triangleq \left [ \mathbf{\Theta}_\mathrm{r}^\mathsf{T}, \mathbf{\Theta}_\mathrm{t}^\mathsf{T} \right ] ^\mathsf{T} \in \mathbb{C}^{2M\times M}$, problem \eqref{eq:Reciprocal_F} can be equivalently transformed into
	\begin{subequations}\label{eq:F_simplified }
		\begin{align}
			\max_\mathbf{F}
			&\sum_{k \in \mathcal{K}}
			\bigg( 
			2\sqrt{1+\iota_k}\Re\left \{\bar{\mathbf{h}}_{k}\mathbf{f}_{k} \right \}-\bigg.
			\mathbf{f}_{k}^\mathsf{H} \sum\limits_{i\in\mathcal{K} } \left(\bar{\mathbf{h}}_{i}^\mathsf{H}  \bar{\mathbf{h}} _{i} \right)\mathbf{f}_{k}
			\bigg ) 	\\
			\text{s.t.}\; &\|\mathbf{F}\|_\mathsf{F}^{2}\leq P_{\mathrm{T}},\label{eq:F_simplified_constraint_1}\\
			&\| \tilde{\mathbf{\Theta}}\mathbf{H}_{\mathrm{IT}} \mathbf{F} \|_\mathsf{F}^2 \leq \bar{P}_{\mathrm{A}},\label{eq:F_simplified_constraint_2}
		\end{align}
	\end{subequations}
	where \(\bar{P}_{\mathrm{A}}\) is defined as
	\begin{equation}
		\bar{P}_{\mathrm{A}}\triangleq\left\{
		\begin{array}{ll}
			P_{\mathrm{A}}-\sigma_{\mathrm{I}}^{2}\left(\left\Vert \mathbf{\Theta}_{\mathrm{r}}\right\Vert _{\mathsf{F}}^{2}+2\left\Vert \mathbf{\Theta}_{\mathrm{t}}\right\Vert _{\mathsf{F}}^{2}\right), &\text{reciprocal},\\
			P_{\mathrm{A}}-\sigma_{\mathrm{I}}^{2}\left(\left\Vert \mathbf{\Theta}_{\mathrm{r}}\right\Vert _{\mathsf{F}}^{2}+\left\Vert \mathbf{\Theta}_{\mathrm{t}}\right\Vert _{\mathsf{F}}^{2}\right), &\text{non-reciprocal}.
		\end{array}\right.
	\end{equation}
	
	Observing \eqref{eq:F_simplified }, the sub-problem regarding \(\mathbf{F}\) is a quadratically constrained quadratic program (QCQP). The Lagrangian multiplier method is adopted to derive the optimal solution to the transmit precoder $\mathbf{F}$. Specifically, we introduce two multipliers $\lambda_1\ge0$ for the constraint \eqref{eq:F_simplified_constraint_1} and  $\lambda_2\ge0$ for the constraint \eqref{eq:F_simplified_constraint_2}, then $\mathbf{F}^{\star}$ for reciprocal and non-reciprocal hybrid mode active BD-RIS can be found uniformly by 
	\begin{equation}\label{eq:F}
		\begin{aligned}
			\mathbf{f}_k^{\star}=&\left( \sum_{i \in \mathcal{K}}\bar{\mathbf{h}}_i^\mathsf{H}\bar{\mathbf{h}}_i
			+\lambda_1^{\star}\mathbf{I}_{N_\mathrm{T}} 
			+\lambda_2^{\star}\mathbf{H}_\mathrm{IT}^\mathsf{H}\tilde{\mathbf{\Theta}}^\mathsf{H}\tilde{\mathbf{\Theta}}\mathbf{H}_\mathrm{IT}
			\right)^{-1}\\
			&~~~~~~~~~~~~~~~~~~~~~\times \sqrt{1+\iota_k}\bar{\mathbf{h}}_k^{\mathsf{H}}, \forall k\in\mathcal{K},
		\end{aligned}
	\end{equation}
	where $\lambda_1^{\star}$ and $\lambda_2^{\star}$ can be calculated via a grid search.
	
	\subsubsection{BD-RIS Matrices $\mathbf{\Theta}_{\mathrm{r}}$ and $\mathbf{\Theta}_{\mathrm{t}}$ }\label{reci_3}
	In this section, we consider the group-connected architecture proposed in Section \ref{sec:architecture} for the optimization of reciprocal and non-reciprocal hybrid mode active BD-RIS, as it provides a general framework including single-connected and fully-connected architectures as special cases.

	\textbf{Reciprocal Active BD-RIS:}
	For the reciprocal case, the constraints for $\mathbf{\Theta}_\mathrm{r}$ and those for $\mathbf{\Theta}_\mathrm{t}$ are different, which motivates us to design $\mathbf{\Theta}_\mathrm{r}$ and $\mathbf{\Theta}_\mathrm{t}$ iteratively. Specifically, when all other variables are fixed, the sub-optimization problem associated with $\mathbf{\Theta}_\mathrm{r}$ is given by
	\begin{subequations}\label{eq:Reciprocal_theta_r_bd}
		\begin{align}
			\max_{\mathbf{\Theta}_{\mathrm r}} \quad
			&\sum_{k\in\mathcal{K}_{\mathrm r}}
			\Big(
			2\sqrt{1+\iota_k}\Re\{\tau_k^* \mathbf{h}_{\mathrm{RI},k}\mathbf{\Theta}_{\mathrm r}\mathbf{H}_{\mathrm{IT}}\mathbf{f}_k\}\notag\\
			&\qquad-|\tau_k|^2\big\|(\mathbf{h}_{\mathrm{RT},k}+\mathbf{h}_{\mathrm{RI},k}\mathbf{\Theta}_{\mathrm r}\mathbf{H}_{\mathrm{IT}})\mathbf{F}\big\|_2^2  \notag\\
			&\qquad\qquad\qquad\quad
			-|\tau_k|^2\sigma_{\mathrm I}^2\big\|\mathbf{h}_{\mathrm{RI},k}\mathbf{\Theta}_{\mathrm r}\big\|_2^2
			\Big), \label{eq:Reciprocal_theta_r_bd_obj}\\
			\text{s.t.}\quad\;
			&\mathbf{\Theta}_{\mathrm r}=\mathbf{\Theta}_{\mathrm r}^{\mathsf T}, \label{eq:Reciprocal_theta_r_bd_sym}\\
			&\mathbf{\Theta}_{\mathrm r}=\mathrm{blkdiag}\!\left(\mathbf{\Theta}_{\mathrm r,1},\ldots,\mathbf{\Theta}_{\mathrm r,\bar G}\right), \label{eq:Reciprocal_theta_r_bd_blk}\\
			&\big\|\mathbf{\Theta}_{\mathrm r}\mathbf{H}_{\mathrm{IT}}\mathbf{F}\big\|_{\mathsf F}^2
			+\sigma_{\mathrm I}^2\big\|\mathbf{\Theta}_{\mathrm r}\big\|_{\mathsf F}^2
			\le P_{\mathrm{A,r}}, \label{eq:Reciprocal_theta_r_bd_pow}
		\end{align}
	\end{subequations}
	where \(P_{\mathrm{A,r}}\triangleq P_{\mathrm A}-(\|\mathbf{\Theta}_{\mathrm t}\mathbf{H}_{\mathrm{IT}}\mathbf{F}\|_{\mathsf F}^2
	+2\sigma_{\mathrm I}^2\big\|\mathbf{\Theta}_{\mathrm t}\big\|_{\mathsf F}^2)\),
	with $\mathbf{\Theta}_{\mathrm t}$ being a constant when designing \(\mathbf{\Theta}_\mathrm{r}\). We further define
	\(\mathbf{C}_{w} \triangleq \sum_{k\in\mathcal{K}_{w}}\sqrt{1+\iota_k}\,\tau_k^*\,\mathbf{H}_{\mathrm{IT}}\mathbf{f}_k\,\mathbf{h}_{\mathrm{RI},k}\), \(\forall w \in \left\{\mathrm{r},\mathrm{t}\right\}\),
	\(\mathbf{D}_{w} \triangleq \sum_{k\in\mathcal{K}_{w}}|\tau_k|^2\,\mathbf{H}_{\mathrm{IT}}\mathbf{F}\mathbf{F}^{\mathsf H}\mathbf{h}_{\mathrm{RT},k}^{\mathsf H}\mathbf{h}_{\mathrm{RI},k}\), \(\forall w \in \left\{\mathrm{r},\mathrm{t}\right\}\),
	\(\mathbf{Q}_{w} \triangleq \sum_{k\in\mathcal{K}_{w}}|\tau_k|^2\,\mathbf{h}_{\mathrm{RI},k}^{\mathsf H}\mathbf{h}_{\mathrm{RI},k}\), \(\forall w \in \left\{\mathrm{r},\mathrm{t}\right\}\), 
	\(\mathbf{P} \triangleq \mathbf{H}_{\mathrm{IT}}\mathbf{F}\mathbf{F}^{\mathsf H}\mathbf{H}_{\mathrm{IT}}^{\mathsf H}+\sigma_{\mathrm I}^2\mathbf{I}_M\), and $\mathbf{E}_{w}\triangleq \mathbf{C}_{w}-\mathbf{D}_{w}$, \(\forall w \in \left\{\mathrm{r},\mathrm{t}\right\}\) where \(\mathbf{C}_{w},\mathbf{D}_{w},\mathbf{E}_{w},\mathbf{Q}_{w},\mathbf{P} \in \mathbb{C}^{M \times M}\). Problem \eqref{eq:Reciprocal_theta_r_bd} can be equivalently rewritten as
	\begin{subequations}\label{eq:Reciprocal_theta_r_1}
		\begin{align}
			\max_{\mathbf{\Theta}_{\mathrm r}}\quad
			&2\Re\{\mathrm{Tr}(\mathbf{E}_{\mathrm r}\mathbf{\Theta}_{\mathrm r})\}
			-\mathrm{Tr}(\mathbf{P}\mathbf{\Theta}_{\mathrm r}^{\mathsf H}\mathbf{Q}_{\mathrm r}\mathbf{\Theta}_{\mathrm r}) \label{eq:Reciprocal_theta_r_1_obj}\\
			\text{s.t.}\quad\;\;
			&\mathbf{\Theta}_{\mathrm r}=\mathbf{\Theta}_{\mathrm r}^{\mathsf T}, \label{eq:Reciprocal_theta_r_1_st1}\\
			&\mathbf{\Theta}_{\mathrm r}=\mathrm{blkdiag}\!\left(\mathbf{\Theta}_{\mathrm r,1},\ldots,\mathbf{\Theta}_{\mathrm r,\bar G}\right), \label{eq:Reciprocal_theta_r_1_st2}\\
			&\mathrm{Tr}(\mathbf{\Theta}_{\mathrm r}\mathbf{P}\mathbf{\Theta}_{\mathrm r}^\mathsf{H})
			\le P_{\mathrm{A,r}}. \label{eq:Reciprocal_theta_r_1_st3}
		\end{align}
	\end{subequations}
	Since \(\mathbf{\mathbf{\Theta}}_{\mathrm r}\) is a block-diagonal matrix, problem \eqref{eq:Reciprocal_theta_r_1} can be further rewritten as 
	\begin{subequations}\label{eq:Reciprocal_theta_r_2}
		\begin{align}
			\max_{\mathbf{\Theta}_{\mathrm r}}\quad
			&2\Re\Big\{\sum_{g=1}^{\bar{G}}\mathrm{Tr}\big(\mathbf{E}_{\mathrm {r},g}\mathbf{\Theta}_{\mathrm {r},g}\big)\Big\}\bigg. \notag\\
			&\bigg.-\mathrm{Tr}\Big(\sum_{g=1}^{\bar{G}}\mathbf{\Theta}_{\mathrm {r},g}^{\mathsf H}\sum_{g'=1}^{\bar{G}}\mathbf{Q}_{\mathrm {r},g,g'}\mathbf{\Theta}_{\mathrm {r},g'}\mathbf{P}_{g',g}\Big) \label{eq:Reciprocal_theta_r_2_obj}\\
			\text{s.t.} \quad\;
			&\mathbf{\Theta}_{\mathrm {r},g}=\mathbf{\Theta}_{\mathrm {r},g}^{\mathsf T}, \forall g\in \mathcal{G}, \label{eq:Reciprocal_theta_r_2_st1}\\
			&\mathrm{Tr}\Big(\sum_{g=1}^{\bar{G}}\mathbf{\Theta}_{\mathrm {r},g}^{\mathsf H}\mathbf{\Theta}_{\mathrm {r},g}\mathbf{P}_{g,g}\Big)	\le P_{\mathrm{A,r}}, \label{eq:Reciprocal_theta_r_2_st2}
		\end{align}
	\end{subequations}
	where \(\mathbf{E}_{\mathrm {r},g}, \mathbf{Q}_{\mathrm {r},g,g'}, \mathbf{P}_{g,g'}\in  \mathbb{C}^{G\times G}\) are respectively defined as \(\mathbf{E}_{\mathrm{r},g} \triangleq [\mathbf{E}_{\mathrm {r}}]_{\mathcal{J}_g,\mathcal{J}_g}\), \(\mathbf{Q}_{\mathrm {r},g,g'}\triangleq [\mathbf{Q}_{\mathrm {r}}]_{\mathcal{J}_g,\mathcal{J}_{g'}}\), and
	\(\mathbf{P}_{g,g'}\triangleq [\mathbf{P}]_{\mathcal{J}_g,\mathcal{J}_{g'}}\) with $\mathcal{J}_g = (g-1)G+1:gG$,
	\( \forall g,g' \in \mathcal{G}\). The constraint \eqref{eq:Reciprocal_theta_r_2_st1} implies that only the diagonal and upper- (or lower-) triangular entries of $\mathbf{\Theta}_{\mathrm{r},g}$, \(\forall g \in \mathcal{G}\), needs to be optimized. This motivates us to identify and extract these entries of interest from $\mathbf{\Theta}_{\mathrm{r},g}$  and reformulate problem \eqref{eq:Reciprocal_theta_r_2} accordingly.
	
	We introduce the vectorization $\bm{\theta}_{\mathrm{r},g} = \mathrm{vec}(\mathbf{\Theta}_{\mathrm{r},g})\in\mathbb{C}^{G^2\times 1}$ and the half-vectorization $\tilde{\bm{\theta}}_{\mathrm{r},g} = \mathrm{vech}(\mathbf{\Theta}_{\mathrm{r},g})\in\mathbb{C}^{\frac{G(G+1)}{2}\times 1}$, which stores the entries of $\mathbf{\Theta}_{\mathrm{r},g}$ below and including the diagonal. Then, we can establish the relationship between half-vectorization and vectorization as
	\begin{equation}
		\bm{\theta}_{\mathrm{r},g}=\mathbf{D}\tilde{\bm{\theta}}_{\mathrm{r},g}
	\end{equation}
	where \(\mathbf{D}\in \mathbb{R}^{ G^2\times \frac{G (G+1)}{2}}\) is a duplication matrix defined as
	%	\begin{equation}
		%		\mathbf{D}^{\mathsf{T}}=\sum_{i \geq j} \mathbf{u}_{ij}\left(\operatorname{vec}\left(\mathbf{T}_{ij} \right)\right)^{\mathsf{T}},
		%	\end{equation}
	\begin{equation}
		\mathbf{D}=\sum_{i \geq j} \operatorname{vec}\left(\mathbf{T}_{ij}\right)\mathbf{u}_{ij}^{\mathsf{T}},
	\end{equation}
	where \(\mathbf{u}_{ij} \in \mathbb{R}^{\frac{G(G+1)}{2}\times 1}\) is a unit vector having value one in  position \((j-1)G+i-\frac{1}{2}j(j-1)\) and zero otherwise, and \(\mathbf{T}_{ij} \in \mathbb{R}^{G\times G}\) is a matrix having value one in positions \(\left(i,j\right)\) and \(\left(j,i\right)\), and zero otherwise. 
	Define \(\bar{\mathbf{Q}}_{\mathrm r} \in \mathbb{C}^{\frac{M(G+1)}{2}\times \frac{M(G+1)}{2}}\) as a block matrix whose \((g,g')\)th block is given by \(	\bar{\mathbf{Q}}_{\mathrm{r},g,g'}
	\triangleq
	\mathbf{D}^{\mathsf T}\!\left(\mathbf{P}_{g',g}^{\mathsf T}\otimes \mathbf{Q}_{\mathrm{r},g,g'}\right)\mathbf{D}
	\in
	\mathbb{C}^{\frac{G(G+1)}{2}\times \frac{G(G+1)}{2}}\),
	\(\bar{\mathbf{P}}_{\mathrm r}\triangleq \operatorname{blkdiag}(\bar{\mathbf{P}}_{1},\ldots, \bar{\mathbf{P}}_{\bar{G}})\in \mathbb{C}^{\frac{M(G+1)}{2}\times \frac{M(G+1)}{2}}\) as a block diagonal matrix whose \(g\)th block is given by
	\(\bar{\mathbf{P}}_{\mathrm{r},g} \triangleq \mathbf{D}^\mathsf{T}(\mathbf{P}_{g}^\mathsf{T}\otimes \mathbf{I}_{G})\mathbf{D} \in \mathbb{C}^{\frac{G(G+1)}{2}\times \frac{G(G+1)}{2}}\), \(\forall g,g' \in \mathcal{G}\),
	\(\bar{\mathbf{e}}_\mathrm{r}\triangleq [\operatorname{vec}(\mathbf{E}_{\mathrm{r},1}^\mathsf{T})^\mathsf{T}\mathbf{D},\ldots,\operatorname{vec}(\mathbf{E}_{\mathrm{r},\bar{G}}^\mathsf{T})^\mathsf{T}\mathbf{D}] \in \mathbb{C}^{1\times\frac{ M(G+1)}{2}}\), and
	\(\bar{\bm{\theta}}_\mathrm{r}=[\tilde{\bm{\theta}}_{\mathrm{r},1}^\mathsf{T},\ldots,\tilde{\bm{\theta}}_{\mathrm{r},\bar{G}}^\mathsf{T}]^\mathsf{T} \in \mathbb{C}^{\frac{ M(G+1)}{2}\times 1}\). We can reformulate problem \eqref{eq:Reciprocal_theta_r_2} through half-vectorization into the following form
	\begin{subequations}\label{eq:Reciprocal_theta_r_vec_sim}
		\begin{align}
			\max_{\bar{\bm{\theta}}_{\mathrm{r}}}\;\;&2\Re\left\{\bar{\mathbf{e}}_\mathrm{r}\bar{\bm{\theta}}_\mathrm{r}\right\}-\bar{\bm{\theta}}_\mathrm{r}^\mathsf{H}\bar{\mathbf{Q}}_\mathrm{r} \bar{\bm{\theta}}_\mathrm{r} \\
			\text{s.t.}\;\;\;\;&\bar{\bm{\theta}}_\mathrm{r}^\mathsf{H}  \bar{\mathbf{P}}_\mathrm{r}\bar{\bm{\theta}}_\mathrm{r} \leq  P_\mathrm{A, r}.\label{eq:Reciprocal_theta_r_vec_sim_con}
		\end{align}
	\end{subequations}
	
	Problem \eqref{eq:Reciprocal_theta_r_vec_sim} is a convex QCQP problem which motivates us to adopt  Lagrangian multiplier method to derive the optimal value.  We introduce a Lagrangian multiplier \(\lambda \ge 0\) for the constraint \eqref{eq:Reciprocal_theta_r_vec_sim_con} and \(\bm{\theta}_\mathrm{r}^\star\) can be calculated from the first-order optimality condition as
	\begin{align}\label{eq:reci_theta_r}
		&\bar{\bm{\theta}}_\mathrm{r}^\star=\left(\bar{\mathbf{Q}}_\mathrm{r}+\lambda^\star\bar{\mathbf{P}}_\mathrm{r}\right)^{-1}
		\bar{\mathbf{e}}_\mathrm{r}^\mathsf{H},
	\end{align}
	where \(\lambda^\star\) can be calculated via a single bisection search. Finally, we can construct \(\bm{\Theta}_\mathrm{r}^\star\) by rearranging entries of \(\bar{\bm{\theta}}_\mathrm{r}^\star\).
	
	The sub-optimization problem associated with \(\mathbf{\Theta}_\mathrm{t}\) is 
	\begin{subequations}\label{eq:Reciprocal_theta_t_bd}
		\begin{align}
			\max_{\mathbf{\Theta}_{\mathrm t}} \quad
			&\sum_{k\in\mathcal{K}_{\mathrm t}}
			\Big(
			2\sqrt{1+\iota_k}\Re\{\tau_k^* \mathbf{h}_{\mathrm{RI},k}\mathbf{\Theta}_{\mathrm t}\mathbf{H}_{\mathrm{IT}}\mathbf{f}_k\}\notag\\
			&\qquad-|\tau_k|^2\big\|(\mathbf{h}_{\mathrm{RT},k}+\mathbf{h}_{\mathrm{RI},k}\mathbf{\Theta}_{\mathrm t}\mathbf{H}_{\mathrm{IT}})\mathbf{F}\big\|_2^2  \notag\\
			&\qquad\qquad\qquad\quad
			-|\tau_k|^2\sigma_{\mathrm I}^2\big\|\mathbf{h}_{\mathrm{RI},k}\mathbf{\Theta}_{\mathrm t}\big\|_2^2
			\Big)\notag\\
			&\qquad\qquad\qquad
			-\sum_{k \in \mathcal{K}_\mathrm{r}}  |\tau_{k}|^2\sigma_{\mathrm{I}}^2\left\|\mathbf{h}_{\mathrm{RI},k} \mathbf{\Theta} _{\mathrm{t}}^\mathsf{T}\right\|_2^2\label{eq:Reciprocal_theta_t_bd_obj}\\
			\text{s.t.}\quad\;\;
			&\mathbf{\Theta}_{\mathrm t}=\mathrm{blkdiag}\!\left(\mathbf{\Theta}_{\mathrm t,1},\ldots,\mathbf{\Theta}_{\mathrm t,\bar G}\right), \label{eq:Reciprocal_theta_t_bd_blk}\\
			&\big\|\mathbf{\Theta}_{\mathrm t}\mathbf{H}_{\mathrm{IT}}\mathbf{F}\big\|_{\mathsf F}^2
			+2\sigma_{\mathrm I}^2\big\|\mathbf{\Theta}_{\mathrm t}\big\|_{\mathsf F}^2
			\le P_{\mathrm{A,t}}, \label{eq:Reciprocal_theta_t_bd_pow}
		\end{align}
	\end{subequations}
	where \(P_{\mathrm{A,t}}\triangleq P_{\mathrm A}-(\|\mathbf{\Theta}_{\mathrm r}\mathbf{H}_{\mathrm{IT}}\mathbf{F}\big\|_{\mathsf F}^2
	+\sigma_{\mathrm I}^2\|\mathbf{\Theta}_{\mathrm r}\|_{\mathsf F}^2)\),
	with $\mathbf{\Theta}_{\mathrm r}$ being a constant when designing \(\mathbf{\Theta}_\mathrm{t}\).	We can transform \eqref{eq:Reciprocal_theta_t_bd} into the following form 
	\begin{subequations}\label{eq:Reciprocal_theta_t_2}
		\begin{align}
			\max_{\mathbf{\Theta}_{\mathrm{t}}} \;\;
			&2\Re \left\{ \mathrm{Tr}\left( \mathbf{E}_{\mathrm{t}}\mathbf{\Theta}_{\mathrm{t}} \right)\right\}
			-\mathrm{Tr}\left(  \mathbf{P}\mathbf{\Theta}_{\mathrm{t}}^\mathsf{H}\mathbf{Q}_{\mathrm{t}}\mathbf{\Theta}_{\mathrm{t}}\right)\bigg.\notag\\
			&\qquad\qquad\qquad\quad\bigg.-\sigma_{\mathrm{I}}^2\mathrm{Tr}\left(  \mathbf{\Theta}_{\mathrm{t}}^*\mathbf{Q}_{\mathrm{r}}\mathbf{\Theta}_{\mathrm{t}}^\mathsf{T}\right)\label{eq:reciprocal_t_2_obj} \\
			\text{s.t.}\quad
			&\mathbf{\Theta}_{\mathrm t}=\mathrm{blkdiag}\!\left(\mathbf{\Theta}_{\mathrm t,1},\ldots,\mathbf{\Theta}_{\mathrm t,\bar G}\right), \label{eq:Reciprocal_theta_t_2_st1}\\
			& \mathrm{Tr}\left(\mathbf{\Theta}_{\mathrm{t}}\left(\mathbf{P} +\sigma_{\mathrm{I}}^2\mathbf{I}_M\right)\mathbf{\Theta}_{\mathrm{t}}^{\mathsf{H}}\right)
			\leq P_\mathrm{A, t}. \label{eq:Reciprocal_theta_t_2_st2} 
		\end{align}
	\end{subequations}
	Since \(\mathbf{\mathbf{\Theta}}_{\mathrm t}\) is a block-diagonal matrix, problem \eqref{eq:Reciprocal_theta_t_2} can be further transformed as 
	\begin{subequations}\label{eq:Reciprocal_theta_t_3}
		\begin{align}
			\max_{\mathbf{\Theta}_{\mathrm t}}\quad
			&2\Re\Big\{\sum_{g=1}^{\bar{G}}\mathrm{Tr}\big(\mathbf{E}_{\mathrm {t},g}\mathbf{\Theta}_{\mathrm {t},g}\big)\Big\}\bigg. \notag\\
			&\bigg.-\mathrm{Tr}\Big(\sum_{g=1}^{\bar{G}}\mathbf{\Theta}_{\mathrm {t},g}^{\mathsf H}\sum_{g'=1}^{\bar{G}}\mathbf{Q}_{\mathrm {t},g,g'}\mathbf{\Theta}_{\mathrm {t},g'}\mathbf{P}_{g',g}\Big)\notag \\
			&\bigg.-\sigma_{\mathrm I}^2\mathrm{Tr}\Big(\sum_{g=1}^{\bar{G}}\mathbf{\Theta}_{\mathrm {t},g}^{\mathsf H}\mathbf{\Theta}_{\mathrm {t},g}\mathbf{Q}_{\mathrm{r},g,g}^\mathsf{T}\Big)\label{eq:Reciprocal_theta_t_3_obj}\\			
			\text{s.t.} \;\;\;\;\;
			&\mathrm{Tr}\big(\sum_{g=1}^{\bar{G}}\mathbf{\Theta}_{\mathrm {t},g}^\mathsf{H}\textstyle\mathbf{\Theta}_{\mathrm {t},g}\left(\mathbf{P}_{g,g}+\sigma_{\mathrm{I}}^2\mathbf{I}_{G}\right)\big)
			\le P_{\mathrm{A,t}}, \label{eq:Reciprocal_theta_t_3_st1}
		\end{align}
	\end{subequations}
	where \(\mathbf{E}_{\mathrm {t},g} \in \mathbb{C}^{G\times G}\) and  \(\mathbf{Q}_{\mathrm {t},g,g'} \in  \mathbb{C}^{G\times G}\) are respectively defined as \(\mathbf{E}_{\mathrm {t},g} \triangleq [\mathbf{E}_{\mathrm {t}}]_{\mathcal{I}_g,\mathcal{I}_g}\) and \(\mathbf{Q}_{\mathrm {t},g,g'}\triangleq [\mathbf{Q}_{\mathrm {t}}]_{\mathcal{I}_g,\mathcal{I}_{g'}}\), \( \forall g,g' \in \mathcal{G}\). To find the optimal solution to all \(\mathbf{\Theta}_{\mathrm {t},g}\) groups simultaneously,
	we define 
	\(\bar{\mathbf{Q}}_{\mathrm t}\in \mathbb{C}^{MG\times MG}\) as a block matrix whose \((g,g')\)th block is given by \(\bar{\mathbf{Q}}_{\mathrm{t},g,g'} \triangleq \mathbf{P}_{ g',g}^\mathsf{T}\otimes \mathbf{Q}_{\mathrm{t},g,g'}+\sigma_{\mathrm{I}}^2\mathbf{Q}_{\mathrm{r}, g,g}\otimes\mathbf{I}_{G} \in \mathbb{C}^{G^2\times G^2}\),
	\(\bar{\mathbf{P}}_{\mathrm t}\triangleq\operatorname{blkdiag}(\bar{\mathbf{P}}_{1},\ldots,\bar{\mathbf{P}}_{\bar{G}})\in \mathbb{C}^{MG\times MG}\) as a block diagonal matrix whose \(g\)th block is given by
	\(\bar{\mathbf{P}}_{\mathrm{t},g} \triangleq \left(\mathbf{P}_{g,g}+\sigma_{\mathrm{I}}^2\mathbf{I}_{G}\right)^\mathsf{T}\otimes \mathbf{I}_{G}\in \mathbb{C}^{G^2\times G^2}\), \(\forall g,g' \in \mathcal{G}\),
	and \(\bar{\mathbf{e}}_\mathrm{t}\triangleq [\operatorname{vec}(\mathbf{E}_{\mathrm{t},1}^\mathsf{T})^\mathsf{T},\ldots,\operatorname{vec}(\mathbf{E}_{\mathrm{t},\bar{G}}^\mathsf{T})^\mathsf{T}] \in \mathbb{C}^{1\times MG}\), 		
	\(\bar{\bm{\theta}}_\mathrm{t}=[\bm{\theta}_{\mathrm{t},1}^\mathsf{T},\ldots,\bm{\theta}_{\mathrm{t},\bar{G}}^\mathsf{T}]^\mathsf{T} \in \mathbb{C}^{MG\times 1}\). Then we reformulate problem \eqref{eq:Reciprocal_theta_t_3} into the following form 
	\begin{subequations}\label{eq:Reciprocal_theta_t_vec_sim}
		\begin{align}
			\max_{\bar{\bm{\theta}}_{\mathrm{t}}}\;\;&2\Re\left\{\bar{\mathbf{e}}_\mathrm{t}\bar{\bm{\theta}}_\mathrm{t}\right\}-\bar{\bm{\theta}}_\mathrm{t}^\mathsf{H}\bar{\mathbf{Q}}_\mathrm{t} \bar{\bm{\theta}}_\mathrm{t} \\
			\text{s.t.}\;\;\; &\bar{\bm{\theta}}_\mathrm{t}^\mathsf{H}  \bar{\mathbf{P}}_\mathrm{t}\bar{\bm{\theta}}_\mathrm{t} \leq  P_\mathrm{A, t}.\label{eq:Reciprocal_theta_t_vec_sim_con}
		\end{align}
	\end{subequations}
	Problem \eqref{eq:Reciprocal_theta_t_vec_sim} is a convex QCQP problem that can be addressed by Lagrangian multiplier method as follows
	\begin{equation}\label{eq:reci_theta_t}
		\bar{\bm{\theta}}_\mathrm{t}^\star=\left(\bar{\mathbf{Q}}_\mathrm{t}+\lambda^\star\bar{\mathbf{P}}_\mathrm{t}\right)^{-1}
		\bar{\mathbf{e}}_\mathrm{t}^\mathsf{H},
	\end{equation} 
	where $\lambda^{\star}$ is a Lagrange multiplier for the constraint \eqref{eq:Reciprocal_theta_t_vec_sim_con} which can be calculated via a bisection search.

	\textbf{Non-Reciprocal Active BD-RIS:} 
	For non-reciprocal case, it is possible to design $\mathbf{\Theta}_{\mathrm{r}}$ and $\mathbf{\Theta}_{\mathrm{t}}$   simultaneously. Specifically, the sub-problem associated with \(\mathbf{\Theta}_{\mathrm{r}}\) and \(\mathbf{\Theta}_{\mathrm{t}}\) is
	\begin{subequations}\label{eq:Nonreciprocal_theta}
		\begin{align}
			\max_{\mathbf{\Theta}_w}
			&\sum_{w \in \left\{\mathrm{t}, \mathrm{r}\right\}}
			\sum_{k \in \mathcal{K}_w}  \bigg(
			2\sqrt{1+\iota_k}\Re\left \{ \tau_k^* \mathbf{h}_{\mathrm{RI},k} \mathbf{\Theta} _{w}\mathbf{H}_{\mathrm{IT}}  \mathbf{f}_{k} \right \} \bigg.\notag\\
			&\qquad\qquad\qquad-|\tau_{k}|^2\left \| \left(\mathbf{h}_{\mathrm{RT},k}+ \mathbf{h}_{\mathrm{RI},k} \mathbf{\Theta} _{w}\mathbf{H}_{\mathrm{IT}}\right)\mathbf{F}  \right \| ^2_2\notag\\
			&\qquad\qquad\qquad\qquad\bigg.-|\tau_{k}|^2\sigma_{\mathrm{I}}^2\left\|\mathbf{h}_{\mathrm{RI},k} \mathbf{\Theta} _{w}\right\|_2^2 \bigg)\\
			\text{s.t.} \;\;
			&\mathbf{\Theta}_{w}=\mathrm{blkdiag}\!\left(\mathbf{\Theta}_{w,1},\ldots,\mathbf{\Theta}_{w,\bar G}\right), \label{eq:NonReciprocal_theta_bd_blk}\\
			&\sum_{w \in \left\{\mathrm{t}, \mathrm{r}\right\}}\left(\left\Vert \mathbf{\Theta}_{w}\mathbf{H}_{\mathrm{IT}}\mathbf{F}\right\Vert _\mathsf{F} ^{2}+\sigma_{\mathrm{I}}^{2}\left\Vert \mathbf{\Theta}_{w}\right\Vert _{\mathsf{F}}^{2}\right)
			\leq P_{\mathrm{A}}.	
		\end{align}
	\end{subequations}
	We define
	\(\tilde{\mathbf{\Theta}}_g\triangleq[\mathbf{\Theta}_{\mathrm{r},g}^\mathsf{T},\mathbf{\Theta}_{\mathrm{t},g}^\mathsf{T}]^\mathsf{T}\in \mathbb{C}^{2G\times G}\),
	\(\mathbf{E}_g\triangleq \left[\mathbf{E}_{\mathrm{r},g}, \mathbf{E}_{\mathrm{t},g}\right]  \in \mathbb{C}^{G\times 2G}\), and 
	\(\mathbf{Q}_{g,g'}\triangleq \mathsf{blkdiag}\left(\mathbf{Q}_{\mathrm{r},{g,g'}}, \mathbf{Q}_{\mathrm{t},{g,g'}}\right) \in\mathbb{C}^{2G\times 2G}\). Problem \eqref{eq:Nonreciprocal_theta} can be reorganized as follows
	\begin{subequations}\label{eq:NonReciprocal_theta_2}
		\begin{align}
			\max_{\tilde{\mathbf{\Theta}}}\quad
			&2\Re\Big\{\sum_{g=1}^{\bar{G}}\mathrm{Tr}\big(\mathbf{E}_{g}\tilde{\mathbf{\Theta}}_{g}\big)\Big\}\bigg. \notag\\
			&\;\bigg.-\mathrm{Tr}\Big(\sum_{g=1}^{\bar{G}}\tilde{\mathbf{\Theta}}_{g}^{\mathsf H}\sum_{g'=1}^{\bar{G}}\mathbf{Q}_{g,g'}\tilde{\mathbf{\Theta}}_{g'}\mathbf{P}_{g',g}\Big) \label{eq:NonReciprocal_theta_2_obj}\\
			\text{s.t.} \;\;\;\;\;&\mathrm{Tr}\Big(\textstyle\sum_{g=1}^{\bar{G}}\tilde{\mathbf{\Theta}}_{g}^{\mathsf H}\tilde{\mathbf{\Theta}}_{g}\mathbf{P}_{g,g}\Big)	\le P_{\mathrm{A}}. \label{eq:NonReciprocal_theta_2_st}
		\end{align}
	\end{subequations}
	Furthermore, we define 
	\(\bar{\mathbf{Q}}\in \mathbb{C}^{2MG\times 2MG}\) as a block matrix whose \((g,g')\)th block is given by \(\bar{\mathbf{Q}}_{g,g'} \triangleq \mathbf{P}_{ g',g}^\mathsf{T}\otimes \mathbf{Q}_{g,g'} \in \mathbb{C}^{2G^2\times 2G^2}\),
	\(\bar{\mathbf{P}}\triangleq \operatorname{blkdiag}(\bar{\mathbf{P}}_{1},\ldots,\bar{\mathbf{P}}_{ \bar{G}})\in \mathbb{C}^{2MG\times 2MG}\), as a block diagonal matrix whose
	\(g\)th block is given by
	\(\bar{\mathbf{P}}_{g} \triangleq \mathbf{P}_{g,g}^\mathsf{T}\otimes \mathbf{I}_{2G}\in \mathbb{C}^{2G^2\times 2G^2}\), \(\forall g,g' \in \mathcal{G}\),
	\(\bar{\mathbf{e}}\triangleq [\operatorname{vec}(\mathbf{E}_{1}^\mathsf{T})^\mathsf{T},\ldots,\operatorname{vec}(\mathbf{E}_{\bar{G}}^\mathsf{T})^\mathsf{T}] \in \mathbb{C}^{1\times 2MG}\), and
	\(\bar{\bm{\theta}}=[\tilde{\bm{\theta}}_{1}^\mathsf{T},\ldots,\tilde{\bm{\theta}}_{\bar{G}}^\mathsf{T}]^\mathsf{T} \in \mathbb{C}^{2MG\times 1}\). Then we reformulate  problem \eqref{eq:NonReciprocal_theta_2} into
	\begin{subequations}\label{eq:Nonreciprocal_theta_vec_sim}
		\begin{align}
			\max_{\bar{\bm{\theta}}}\;\;&2\Re\left\{\bar{\mathbf{e}}\;\bar{\bm{\theta}}\right\}-\bar{\bm{\theta}}^\mathsf{H}\bar{\mathbf{Q}} \;\bar{\bm{\theta}} \\
			\text{s.t.}\quad &\bar{\bm{\theta}}^\mathsf{H}\bar{\mathbf{P}} \;\bar{\bm{\theta}} \leq  P_\mathrm{A},\label{eq:Nonreciprocal_theta_vec_sim_con}
		\end{align}
	\end{subequations}		
	that is again a convex QCQP and can be solved by the Lagrange multiplier method in closed form as
	\begin{equation}\label{eq:nonreci_theta}
		\bar{\bm{\theta}}^\star=\left(\bar{\mathbf{Q}}+\lambda^\star\bar{\mathbf{P}}\right)^{-1}
		\bar{\mathbf{e}}^\mathsf{H},
	\end{equation} 
	where $\lambda^{\star}$ is a Lagrange multiplier for the constraint \eqref{eq:Nonreciprocal_theta_vec_sim_con} and can be calculated via a bisection search.
	%\subsection{Convergence Analysis}
	
	    \begin{table*}[]
		\centering
        \begin{threeparttable}
            \caption{Complexity Comparison of Different Active BD-RIS Architectures}\label{tab:complexity}
		\begin{tabular}{|c|c|c|c|c|}
			\hline
			Cell-Wise Architecture                     & No. of Groups & Group Size         & \makecell{No. of \\ Non-zero Elements}    & Optimization Complexity\tnote{\(\dagger\)}  \\ \hline
			Single-Connected                 & $M$                                                          & \(1\)                   & $2M$                                                                     & \(\mathcal{O}\left\{I\left(C+I_{\mathsf{bs},3}M^3 \right)\right\}\)                        \\ \hline
			\multirow{2}{*}{Group-Connected} & \multirow{2}{*}{$\bar{G}$}                                     & \multirow{2}{*}{\(G\)} & \multirow{2}{*}{\(2MG\)}                                                   & \(\mathcal{O}\left\{I\left(C+M^{2}G^{4}+\left(I_{\mathsf{bs},1}+I_{\mathsf{bs},2}\right)M^{3}G^{3}\right)\right\}
			\) for reciprocal case    \\ \cline{5-5} 
			&                                                             &                    &                                                                        & \(\mathcal{O}\left\{I\left(C+I_{\mathsf{bs},3}M^3G^3 \right)\right\}\) for non-reciprocal case  \\ \hline
			\multirow{2}{*}{Fully-Connected} & \multirow{2}{*}{\(1\)}                                          & \multirow{2}{*}{\(M\)} & \multirow{2}{*}{\(2M^2\)}                                 & \(\mathcal{O}\left\{I\left(C+\left(I_{\mathsf{bs},1}+I_{\mathsf{bs},2}\right)M^{6}\right)\right\}
			\) for reciprocal case     \\ \cline{5-5} 
			&                                                             &                    &                                                                        & \(\mathcal{O}\left\{I\left(C+I_{\mathsf{bs},3}M^6 \right)\right\}\) for non-reciprocal case \\ \hline
		\end{tabular}
        \begin{tablenotes}
            \footnotesize
            \item[\(\dagger\)] \(I\) denotes the number of iterations for the proposed unified optimization algorithm.
        \end{tablenotes}
        \end{threeparttable}
		
	\end{table*}

	\subsection{Computational Complexity Analysis}
	This section briefly analyzes the computational complexity of the proposed unified optimization framework for hybrid mode active BD-RIS. As illustrated in Algorithm~\ref{alg:joint_precoder_bdris}, four blocks are optimized iteratively. Specifically, the updates of the two auxiliary variables \(\bm{\tau}\) and \(\bm{\iota}\) incur a per-iteration complexity of \(\mathcal{O}\!\left(K^{2}M^{2}\right)\).
	The optimization of the transmit precoder \(\mathbf{F}\) has a complexity of 
	\(\mathcal{O}\!\left(K\!\left(I_\mathsf{grid} N_\mathrm{T}^3 + N_\mathrm{T}M^2\right)\right)\), where $I_\mathsf{grid}$ denotes the number of grid searches associated with constraints \eqref{eq:F_simplified_constraint_1} and \eqref{eq:F_simplified_constraint_2}.
	For the group-connected hybrid mode active BD-RIS with group size $G$, the optimization complexity for reciprocal and non-reciprocal architectures are respectively \(\mathcal{O}\!\left(M^{2}G^{4}+\left(I_{\mathsf{bs},1}+I_{\mathsf{bs},2}\right)M^{3}G^{3}\right)\) and \(\mathcal{O}\!\left(I_{\mathsf{bs},3}M^{3}G^{3}\right)\), where \(I_{\mathsf{bs},1}\), \(I_{\mathsf{bs},2}\), and \(I_{\mathsf{bs},3}\) denote the numbers of bisection searches associated with constraints \eqref{eq:Reciprocal_theta_r_vec_sim_con}, \eqref{eq:Reciprocal_theta_t_vec_sim_con} and \eqref{eq:Nonreciprocal_theta_vec_sim_con}, respectively.
	For clarity, we define a common complexity term \(C \triangleq K^{2}M^{2} +K\!\left(I_\mathsf{grid} N_\mathrm{T}^3 + N_\mathrm{T}M^2\right)\) shared by all hybrid mode active BD-RIS architectures. A detailed comparison of the complexity associated with different hybrid mode active BD-RIS architectures is provided in Table~\ref{tab:complexity}.
	
	%Since different active BD-RIS architectures require distinct impedance components, we further conduct a systematic analysis of the number of non-zero elements and the associated circuit topology  to gain deeper insights into the complexity--performance tradeoff. As shown in Table~\ref{tab:complexity}, increasing the interconnection degree from single-connected to fully-connected architecture substantially enhances the structural flexibility, at the cost of a rapidly growing circuit topology and optimization complexity. In particular, the group-connected architecture offers a favorable tradeoff by maintaining scalable hardware complexity while effectively avoiding the prohibitive \(\mathcal{O}(M^{6})\) optimization burden inherent to fully-connected designs.

	\section{Performance Evaluation}
	\label{sec:evaluation}
	\begin{figure}[!t]
		\centering
		\includegraphics[width=0.48\textwidth]{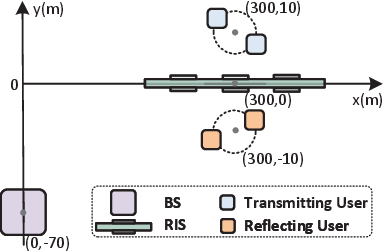}
		\caption{An illustration of the relative position among the BS, active BD-RIS, and users based on a 2D coordinate.}
		\label{fig:topview}
	\end{figure}

	In this section, we evaluate the sum rate performance of the MU-MISO system aided by a hybrid mode active BD-RIS. Simulation results in this section are based on the following parameters. All the channels have large-scale fading characterized by \(\mathrm{PL}|_{\mathrm{dB}}=41.2+28.7\log d\) based on the 3GPP standard \cite{prevail2023}, where \(d\) is the distance between devices, and small-scale fading characterized by the Rician fading model with the Rician factor \(\kappa=1\). That is, we assume that there exist strong line-of-sight links between the BS and active BD-RIS and between the active BD-RIS and users. 
	The BS, active BD-RIS, and users are located in a 2D coordinate system as illustrated in Fig. \ref{fig:topview}. Specifically, the BS and active BD-RIS 
	are located at [0, -70m] and [300m, 0], respectively.
	The transmitting and reflecting users are randomly distributed within circular regions of radius 3 meters centered at [300m, 10m] and [300m, -10m], respectively.
	The noise powers at active BD-RIS and all users are set to \(\sigma_\mathrm{I}^2=\sigma_\mathrm{R}^2=-90\) dBm.
	For fair comparison, the total transmit power budget of a wireless communication system without RIS is denoted by \(P_\mathrm{T}^{\mathrm{tot}}\). 
	For the hybrid mode active BD-RIS aided system, the transmit power and the amplification power at the RIS are allocated as 
	$P_\mathrm{T} = 0.99P_\mathrm{T}^{\mathrm{tot}}$ and $P_\mathrm{A} = 0.01P_\mathrm{T}^{\mathrm{tot}}$, respectively. 
	In contrast, for the system aided by passive BD-RIS with hybrid mode, the entire power budget is used for transmission, i.e., $P_\mathrm{T} = P_\mathrm{T}^{\mathrm{tot}}$.
    
    In the following simulations, we include five schemes as summarized below. 
    \begin{enumerate}
        \item \textit{Hybrid Mode Active BD-RIS} with reciprocal and non-reciprocal cell-wise fully-connected (CWFC) and cell-wise group-connected (CWGC) architectures as summarized in Table \ref{tab:architecture}.
        \item \textit{Active STAR-RIS} \cite{xu2023,Maghrebi2024,Faramarzi2025,wang2026,aung2025,rongguang2025,huang2025} that is a special case of reciprocal hybrid mode active BD-RIS with cell-wise single-connected architecture.
        \item \textit{Hybrid Mode Passive BD-RIS} with non-reciprocal CWFC and CWGC architectures proposed in \cite{first}. 
        In this case, the beamforming design is obtained by reusing the fractional programming method \cite{FPI}. Specifically for the design of hybrid mode passive BD-RIS, the constraint \(\mathbf{\Theta}_\mathsf{r}^{\mathsf{H}}\mathbf{\Theta}_\mathsf{r} + \mathbf{\Theta}_\mathsf{t}^{\mathsf{H}}\mathbf{\Theta}_\mathsf{t}=\mathbf{I}_{M}\) is considered and addressed by the conjugate-gradient method on the manifold of unitary matrices as proposed in \cite{first}. Since hybrid mode passive BD-RISs with non-reciprocal architectures provide more flexibility in wave manipulation than those with reciprocal architectures and thus achieve higher performance, here we do not include the results for hybrid mode passive BD-RIS with reciprocal architectures without loss of observations. 
        \item \textit{Passive STAR-RIS} that is a special case of hybrid mode passive BD-RIS with cell-wise single-connected architecture.
         \item \textit{w/o RIS} that refers to the scenario without the deployment of RISs between the transceivers.
    \end{enumerate}

	\begin{figure}[!t]
		\centering
		\subfigure[\(N_\mathrm{T}=4\), \(K_\mathrm{r}=K_\mathrm{t}=2\), \(M=16\)]{
			\includegraphics[width=0.48\textwidth]{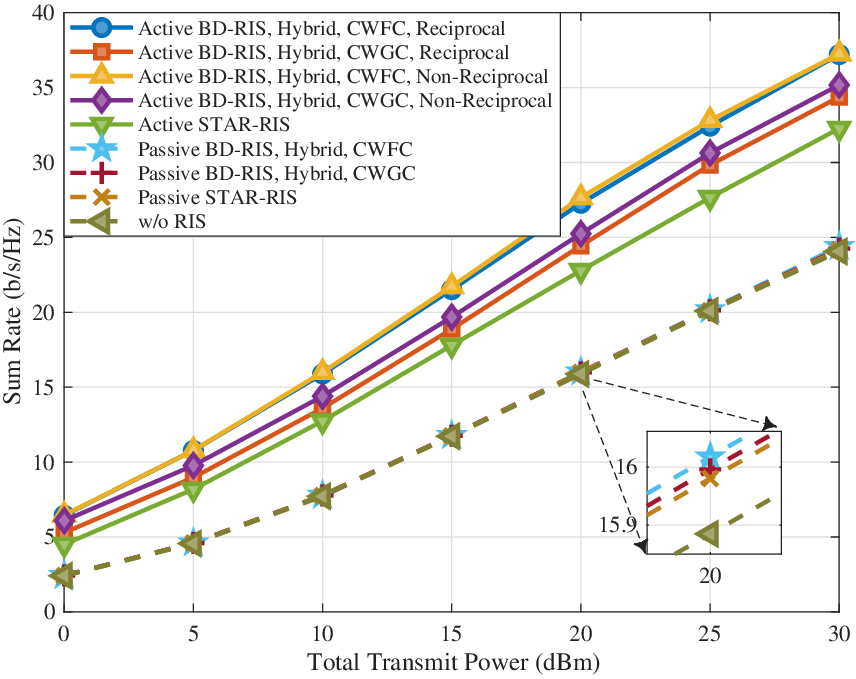}}
		\subfigure[\(N_\mathrm{T}=6\), \(K_\mathrm{r}=K_\mathrm{t}=3\), \(M=24\)]{
			\includegraphics[width=0.48\textwidth]{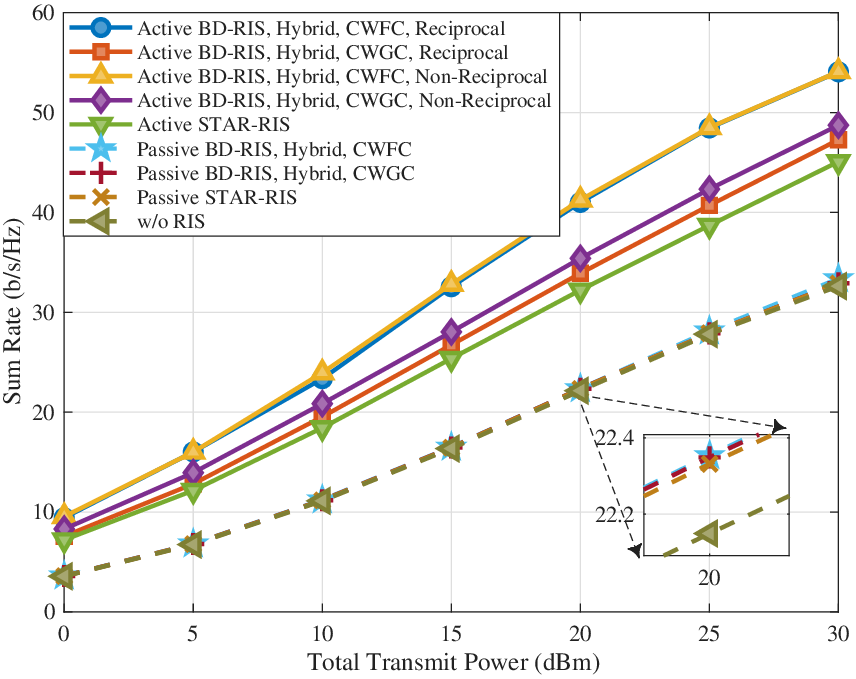}}
		\caption{Sum rate versus transmit power \(P_\mathrm{T}^{\mathrm{tot}}\). \(G=2\) for group-connected architecture.}\vspace{-0.2 cm}
		\label{fig:sr_vs_p}
	\end{figure}

		\begin{figure}[!t]
		\centering
		\subfigure[\(N_\mathrm{T}=4\), \(K_\mathrm{r}=K_\mathrm{t}=2\)]{
			\includegraphics[width=0.48\textwidth]{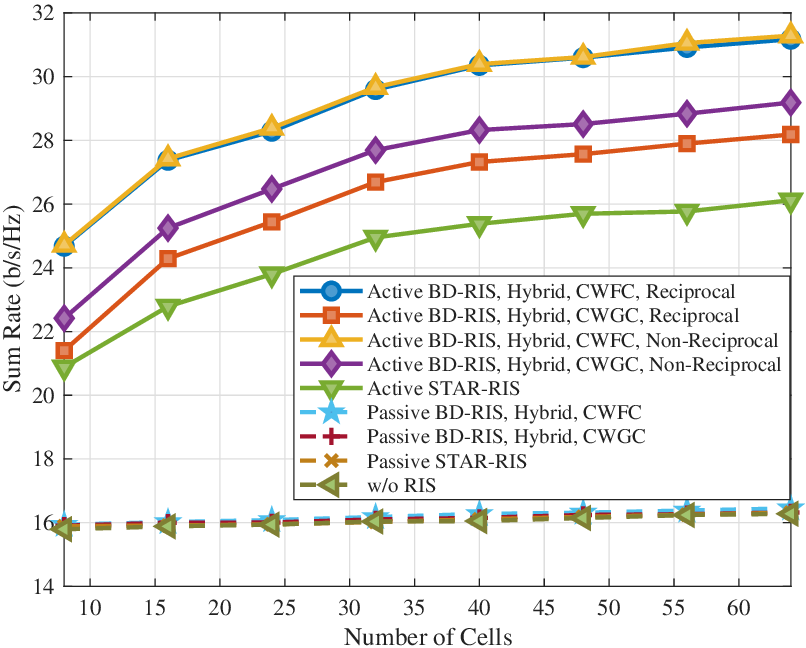}}
		\subfigure[\(N_\mathrm{T}=6\), \(K_\mathrm{r}=K_\mathrm{t}=3\)]{
			\includegraphics[width=0.48\textwidth]{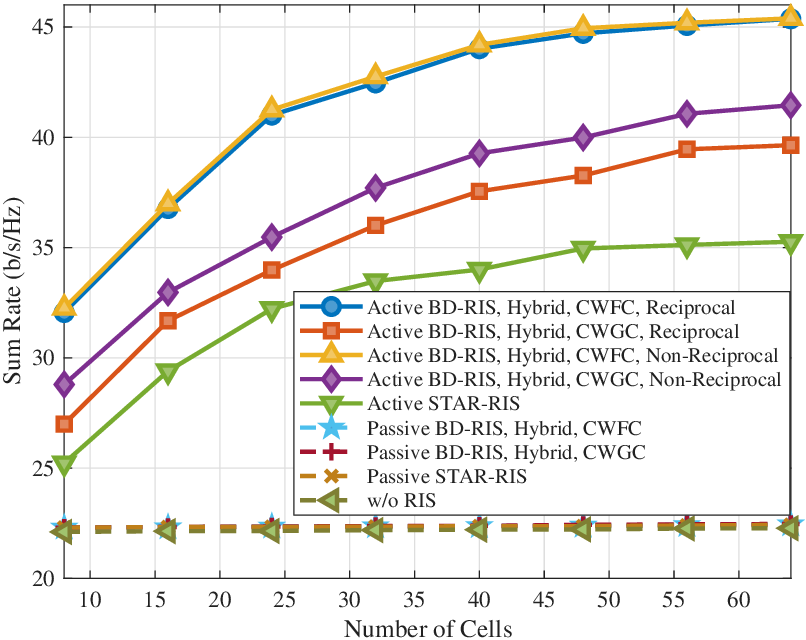}}
            \vspace{-1mm}
		\caption{Sum rate versus the number of RIS cells \(M\). \(P_\mathrm{T}^{\mathrm{tot}}\)=20 dBm.}\vspace{-0.3 cm}
		\label{fig:sr_vs_e}
	\end{figure}
	
	In Fig.~\ref{fig:sr_vs_p}, we present the sum rate performance of an MU-MISO communication system aided by an active BD-RIS with hybrid mode versus the total transmit power \(P_\mathrm{T}^{\mathrm{tot}}\).
	From Fig.~\ref{fig:sr_vs_p}, we can obtain the following observations.
	\textit{First}, 
	active BD-RISs provide significant performance improvements over passive BD-RISs when a strong direct link exists between the transceivers.
	 For instance,
	in an MU-MISO system with 4 transmit antennas and a 16-cell active BD-RIS, active BD-RIS with non-reciprocal CWFC architecture improves the sum rate by 54.8\% at a transmit power of 30 dBm, whereas the passive counterpart only achieves an improvement of 1.4\%. This notable performance gap indicates that hybrid mode active BD-RIS can effectively compensate for the multiplicative fading by amplifying scattered signals, thereby providing substantial performance improvements compared to the passive counterpart.
	\textit{Second},
	active BD-RIS with CWGC/CWFC architecture consistently outperforms active STAR-RIS, and the performance advantage becomes more evident with more BD-RIS cells.
	 For instance,
	 in an MU-MISO system with 6 transmit antennas and a 24-cell active BD-RIS, hybrid mode active BD-RIS with non-reciprocal CWFC and CWGC architectures achieve sum-rate improvements of approximately 20\% and 8\%, respectively, compared with an active STAR-RIS of the same size at a transmit power of 30 dBm.
   Meanwhile,  
     active BD-RIS with non-reciprocal CWFC architecture and passive counterpart improves the sum rate by approximately 65.8\% and 2.2\% at a transmit power of 30 dBm.
	This improvement mainly arises from the intrinsic inter-element connections in BD-RIS architectures, which allow more flexible manipulation of the scattered signals at the cost of higher circuit complexity.
	\textit{Third},
	the performance difference between reciprocal and non-reciprocal active BD-RIS architectures is relatively small under the considered simulation settings. In particular, the reciprocal and non-reciprocal CWFC architectures achieve nearly identical sum rates. Furthermore, in an MU-MISO system with 6 transmit antennas, the non-reciprocal CWGC architecture (with a group size of 2) improves the performance by only about 3\% compared with its reciprocal counterpart. Given the higher circuit complexity associated with non-reciprocal implementations, these results suggest that reciprocal active BD-RIS architectures may offer a more favorable trade-off between achievable performance and hardware complexity. 
	
	\begin{figure}[!t]
		\centering
		\subfigure[\(N_\mathrm{T}=4\), \(K_\mathrm{r}=K_\mathrm{t}=2\), \(M=16\)]{
			\includegraphics[width=0.48\textwidth]{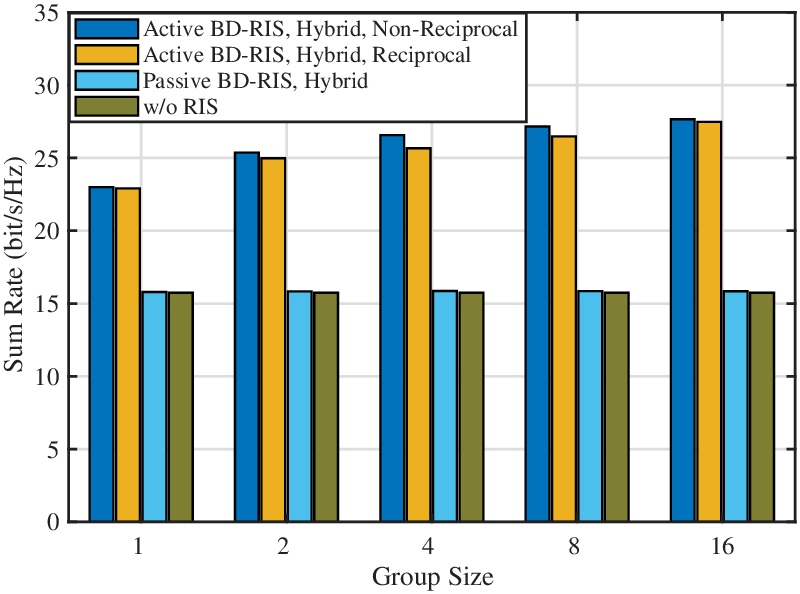}\label{fig:sr_vs_g1}}
		\subfigure[\(N_\mathrm{T}=6\), \(K_\mathrm{r}=K_\mathrm{t}=3\), \(M=24\)]{
			\includegraphics[width=0.48\textwidth]{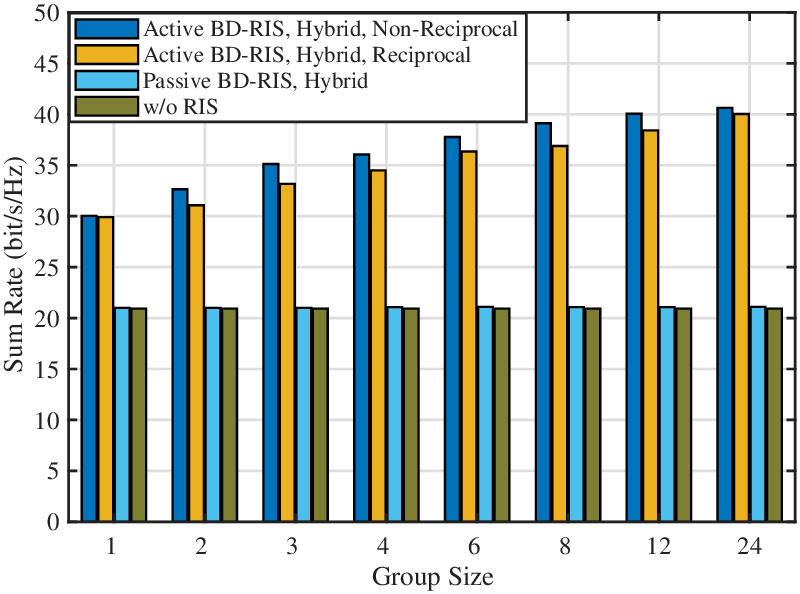}\label{fig:sr_vs_g2}}
		\caption{Sum rate versus the group size \(G\) of active BD-RIS. \(P_\mathrm{T}^{\mathrm{tot}}\)=20 dBm.}\vspace{-0.3 cm}
		\label{fig:sr_vs_g}
	\end{figure}
	
	In Fig. \ref{fig:sr_vs_e}, we plot the sum rate of an MU-MISO system versus the number of active BD-RIS cells with \(P_\mathrm{T}^{\mathrm{tot}}=20\) dBm.
	We can have two main observations from Fig. \ref{fig:sr_vs_e}.
	 \textit{First}, 
     the sum rate improves as the number of BD-RIS cells increases for all active architectures, whereas the performance of passive BD-RIS and the case without RIS remains nearly unchanged.
	 This is because increasing the number of active BD-RIS cells simultaneously involves more reflection-type amplifiers, which enhance the beamforming capability and improve the effective channel gain. 
	 In contrast, the performance gain of passive BD-RISs is limited by multiplicative fading, especially when there exist strong line-of-sight links between transceivers.
	 \textit{Second}, the proposed active BD-RIS consistently outperforms the active STAR-RIS across different architectures, and the superiority of the proposed scheme becomes increasingly evident as the number of cells grows. This gain can be attributed to the extra degrees of freedom enabled by the inter-element connections in hybrid mode active BD-RIS architectures, which allow more flexible manipulation of the scattered and amplified signals. 
	 Overall, these results demonstrate that active BD-RIS can effectively leverage an increasing number of elements to achieve substantial performance gains, highlighting its superiority over active and passive STAR-RIS as well as passive BD-RIS with hybrid mode.

	In Fig. \ref{fig:sr_vs_g}, we present the sum rate performance of an MU-MISO communication system versus the group size of the active BD-RIS with \(P_\mathrm{T}^{\mathrm{tot}}=20\) dBm.
	The group size \(G\) is selected from the set of divisors of the number of active BD-RIS cells. In particular, \(G=1\) and \(G=M\) correspond to the cell-wise single-connected (STAR-RIS) and CWFC architectures, respectively.
	From Fig. \ref{fig:sr_vs_g}, we can obtain the following observations.
	\textit{First}, 
	the achievable sum rate increases with the group size for both reciprocal and non-reciprocal active BD-RIS architectures. This trend indicates that larger group sizes can provide greater flexibility in controlling the scattered signals, thus improving system performance.
	\textit{Second},
	the non-reciprocal active BD-RIS consistently achieves a slightly higher sum rate than the reciprocal architecture for different group sizes, which can be attributed to the extra degrees of freedom enabled by the non-reciprocal implementation.

	\section{Conclusion}
	\label{sec:conclusion}
	
	In this paper, we investigate an active BD-RIS with hybrid transmitting and reflecting mode from the perspectives of modeling, architecture design, and optimization.
    An \(M\)-cell hybrid mode active BD-RIS is characterized as a \(2M\)-port active reconfigurable impedance network connected to $2M$ antenna elements. 
    The $2M$-port active reconfigurable impedance network is constructed by a $4M$-port passive reconfigurable impedance network whose half ports are connected to $2M$ amplifiers.
     Based on the circuit topology of the passive reconfigurable impedance network, we analyze the reciprocal and non-reciprocal characteristics and propose three inter-cell connection architectures, namely cell-wise single, group, and fully-connected architectures. Furthermore, we prove that active STAR-RIS and active double-faced RIS are essentially special cases of the proposed model, i.e. reciprocal hybrid mode active BD-RIS with cell-wise single-connected architecture.
	
	We then jointly optimize the transmit precoding and hybrid mode active BD-RIS for an MU-MISO system with the objective of sum rate maximization. To address this problem, a unified optimization framework is proposed that is applicable to all architectures considered. 
		Simulation results demonstrate that, for an MU-MISO system with 4 transmit antennas and a 16-cell active BD-RIS, even in the presence of strong line-of-sight links between transceivers, the hybrid mode active BD-RIS with non-reciprocal CWFC architecture can achieve a sum rate improvement by 54.8\%, whereas active STAR-RIS and passive BD-RIS with non-reciprocal CWFC architecture achieve an improvement of about 34.1\% and 1.4\% compared to the case without RIS, respectively.

		\bibliographystyle{IEEEtran}
		\bibliography{references}
	\end{document}